%% file: izs24PMC_v4.tex
\documentclass[a4paper,conference]{IEEEtran}

\usepackage{ifthen} 
\newboolean{short_version}
\setboolean{short_version}{false} 

\IEEEsettopmargin{t}{30mm}
\IEEEquantizetextheight{c}
\IEEEsettextwidth{14mm}{14mm}
\IEEEsetsidemargin{c}{0mm}

\usepackage[utf8]{inputenc}
\usepackage[T1]{fontenc}
\usepackage{balance}
\usepackage{ifthen}
\usepackage[cmex10]{amsmath} 
\usepackage[lined,boxed,commentsnumbered,linesnumbered,ruled]{algorithm2e}
\usepackage[url,hyperrefblack,notheorems,IEEEtran]{research17} 
\usepackage{amsmath,amssymb,amsfonts,amsbsy}
\usepackage{afterpage,refcount}
\newcommand{\setfootnotemark}{%
  \refstepcounter{footnote}%
  \footnotemark[\value{footnote}]}
\usepackage{booktabs}
\usepackage{multirow}
\usepackage{cellspace}
\usepackage{enumitem}
\usepackage[capitalize]{cleveref}

\usepackage{xargs} 
\usepackage[colorinlistoftodos, prependcaption, textsize=footnotesize]{todonotes} 
\newcommandx{\unsure}[2][1=]{\todo[linecolor=red,backgroundcolor=red!25,bordercolor=red,#1]{#2}}
\newcommandx{\change}[2][1=]{\todo[inline, inlinewidth=0.95\linewidth, linecolor=blue,backgroundcolor=blue!25,bordercolor=blue,#1]{#2}}
\newcommandx{\info}[2][1=]{\todo[inline, inlinewidth=0.9\linewidth,linecolor=niceGreen,backgroundcolor=niceGreen!25,bordercolor=niceGreen,#1]{#2}}
\newcommandx{\improvement}[2][1=]{\todo[linecolor=purple,backgroundcolor=purple!25,bordercolor=purple,#1]{#2}}

\DeclareMathOperator{\diam}{diam}
\DeclareMathOperator{\modstar}{mod}

\newtheorem{definition}{\mydefinitionname}
\newtheorem{remark}{\myremarkname}
\newtheorem{example}{\myexamplename}
\newcommand{\collect}[1]{\mathscr{#1}} 
\newcommand*{\Resize}[2][4]{\resizebox{#1}{!}{\ensuremath{#2}}} 
\renewcommand{\r}{\color{red}}
\renewcommand{\b}{\color{blue}} 
\definecolor{lightBlue}{RGB}{74,144,226} 
\definecolor{darkRed}{RGB}{208,2,27} 
\definecolor{darkYellow}{RGB}{245,166,35}
\definecolor{niceGreen}{RGB}{126,211,33}
\definecolor{purple}{RGB}{189,16,224}

\definecolor{color2}{RGB}{255,192,203}
\definecolor{color3}{RGB}{255,240,100}
\definecolor{color5}{RGB}{229,204,255}

\begin{document}

\title{Improved Capacity Outer Bound for Private Quadratic Monomial Computation}
\author{%
    \IEEEauthorblockN{Karen M.~D{\ae}hli, Sarah A.~Obead, Hsuan-Yin Lin, and Eirik Rosnes}
    \IEEEauthorblockA{Simula UiB, N--5006 Bergen, Norway}
    \IEEEauthorblockA{Emails: kamadaehli@gmail.com, \{sarah, lin, eirikrosnes\}{@}simula.no  }
}
\maketitle

\begin{abstract}
In private computation, a user wishes to retrieve a function evaluation of messages stored on a set of databases without revealing the function's identity to the databases. 
Obead \emph{et al.} introduced a capacity outer bound for private nonlinear computation, dependent on the order of the candidate functions. 
Focusing on private \emph{quadratic monomial} computation, we propose three methods for ordering candidate functions: a graph edge-coloring method, a graph-distance method, and an entropy-based greedy method. We confirm, via an exhaustive search, that all three methods yield an optimal ordering for $f < 6$ messages. For $6 \leq f \leq 12$ messages, we numerically evaluate the performance of the proposed methods compared with a directed random search.  For almost all scenarios considered,  the entropy-based greedy method gives the smallest gap to the best-found  ordering. 
\end{abstract}

\section{Introduction}
\label{sec:introduction}
\vspace{-0.5ex}

Private computation (PC) \cite{SunJafar19_2} is a generalization of the renowned private information retrieval (PIR) problem that aims at addressing privacy concerns in distributed computing services.
For example, in distributed machine learning, many common classification, dimensionality reduction, and linear regression algorithms operate on the inner products of the data samples rather than the individual data samples. In PC, the user wants to privately download a function evaluation of the messages stored across a set of databases, i.e., without leaking any information to the databases (in an information-theoretic manner) on the identity of the desired function evaluation. 

To measure the efficiency of a PC protocol, the PC rate, defined as the ratio between the (smallest) size of the function evaluation and the number of downloaded symbols, is typically considered. The maximum PC rate is referred to as the PC capacity, and it is known for the case of linear function evaluations, referred to as private linear computation (PLC), from noncolluding replicated and coded databases \cite{SunJafar19_2, ObeadKliewer18_1, ObeadLinRosnesKliewer22_1}. 
Private polynomial computation (PPC) was first considered by Karpuk in \cite{Karpuk18_1}, and later in \cite{RavivKarpuk19_2, ObeadLinRosnesKliewer22_2, YakimenkaLinRosnes20_1, ZhuYanTangLi22_1}. The capacity of PPC is still unknown, and there is generally a substantial gap between the best achievable rate and the best-known capacity outer bound. Private inner product retrieval from noncolluding replicated databases was considered in \cite{MousaviMaddah-AliMirmohseni19_1}, while the general case of PC for nonlinear function evaluations was considered in \cite{ObeadLinRosnesKliewer19_2} where an outer bound on the capacity was first introduced.  
It was noted in \cite{ObeadLinRosnesKliewer19_2} 
that the value of the PC capacity outer bound depends on how the candidate functions are ordered.\footnote{Due to the inherent relation between PC and PIR, a similar observation was made in \cite{ChenWangJafar20_1} for  PIR with dependent messages, i.e., dependent PIR (DPIR).} %
To the best of our knowledge, an optimal order for the outer bound has not yet been considered in the open literature.

In this work, inspired by private inner product retrieval \cite{MousaviMaddah-AliMirmohseni19_1}, we focus on the case of private quadratic nonparallel monomial computation (PQNMC) and propose three methods for finding a \emph{good} ordering of the $\mu$ candidate functions. 
In PQNMC, the set of candidate functions is the set of all quadratic \emph{nonparallel} monomials of $f$ messages where each message symbol is chosen from a size-$q$ finite field $\Field_q$, thus, $\mu=\nicefrac{f(f-1)}{2}$. %
Given that graphs present a strong framework for illustrating the interdependence among random variables, offering insights into the dependency structure of the candidate function set, 
we propose two graph-based methods. The first is an edge-coloring method we name \emph{(enhanced) edge-coloring} ((E-)EC) and the second is a graph-distance method we name \emph{longest-distance first} (LDF). Then, we compare the resulting PC capacity outer bound with the one found by our third method: an \emph{entropy-based greedy} (EBG) algorithm.

For $f < 6$ messages chosen from $\Field_2$, we verify through an exhaustive search that the proposed methods output \emph{optimal} orders.
However, we note that the orders are not unique and are finite field-dependent, which illustrates the difficulty of finding a general ordering of quadratic nonparallel monomials that will optimize the outer bound of the PQNMC capacity. %
Moreover, for larger numbers of messages, an exhaustive search quickly becomes infeasible even over $\Field_2$. As a result, we opt for a \emph{directed} random search to numerically analyze the performance of the proposed methods. Accordingly, we note that for $6 \leq f\leq 10$ and $f=12$ messages, the EBG algorithm outperforms the proposed graph-based methods with the smallest gap to the best-found ordering.  %
Nevertheless, the significance of the graph-based methods arises as a relatively low-complexity alternative to the EBG method as the complexity of computing the entropies needed for the EBG algorithm grows exponentially with the number of candidate monomials $\mu$ with base equal to the size $q$ of the underlying finite field $\Field_q$.
Finally, although we consider PQNMC in this work, we note that the results may also have independent interest beyond PC.\looseness-1

\section{Preliminaries}
\label{sec:preliminaries}

\subsection{Notation}
\label{sec:notation}

We denote by $\Naturals$ the set of all positive integers and $[a]\eqdef\{1,2,\ldots,a\}$ for $a\in \Naturals$. %
A random variable is denoted by a capital Roman letter, e.g., $X$, while its realization is denoted by the corresponding small Roman letter, e.g., $x$. 
Vectors are boldfaced, e.g., $\vect{X}$ denotes a random vector, and $\vect{x}$ denotes a deterministic vector. %
Sets are denoted by calligraphic uppercase letters, e.g., $\set{X}$. Concatenation of vectors $\vect{x}_1, \dots, \vect{x}_a$ is represented by $(\vect{x}_1\mid\cdots\mid\vect{x}_a)$. %
Furthermore, some constants and functions are depicted by Greek letters or a special font, e.g., $\const{X}$. The entropy of $X$ is represented by $\HP{X}$. %
A degree $g$ monomial ${\vect{z}}^{\vect{i}}$ in $f$ variables $z_1,\ldots,z_f$ over a finite field $\Field_q$ is written as ${\vect{z}}^{\vect{i}} = z^{i_1}_1 \cdots z^{i_f}_f$, where $\vect{i}\eqdef(i_1,\ldots,i_f)\in (\{0\} \cup \Naturals)^{f}$ is the exponent vector with $\sum_{j=1}^{f}i_j = g$ and $i_j\leq q-1$ for all $j\in [f]$.\footnote{For nonvanishing polynomials, following the Combinatorial Nullstellensatz theorem \cite[Thm.~1.2]{Alon99_1}, the degree of every variable in a multivariate polynomial must  be strictly smaller than the finite field size.}
A simple undirected graph with vertex set $\set{V}$ and edge set $\set{E}$ is denoted by $\set{G}=(\set{V},\set{E})$.\looseness-1

\subsection{Problem Statement}
\label{sec:problem-statement}
We consider the PQNMC problem, which is formally described as follows.\footnote{A monomial $m(\vect{z})$ is said to be \emph{parallel} if it can be raised by another monomial to a positive integer power, i.e., $m(\vect{z})=(\vect{z}^{\vect{i}})^d$ for some $d\in\Naturals$ and $\vect{i}\in (\{0\} \cup \Naturals)^{f}$.}
Consider a distributed storage system (DSS) consisting of $n$ noncolluding databases, each storing a replica of $f$ independent messages.  The messages are denoted by $\vect{W}^{(1)},\ldots,\vect{W}^{(f)}$ and each message $\vect{W}^{(m)}=\bigl(W_{1}^{(m)},\dots, W_{\beta\const{L}}^{(m)}\bigr)$, $m\in [f]$, is a length-$\beta\const{L}$ vector with independent and identically distributed symbols that are chosen uniformly at random from the field $\Field_q$ for some $\beta,\const{L}\in\Naturals$.\footnote{%
For consistency, we use the notation required for the achievable rate in \cite[Thm.~2]{ObeadLinRosnesKliewer19_2} where asymptotically $\const{L}\rightarrow\infty$ but $\beta$ is fixed.} Hence, we have
\begin{IEEEeqnarray*}{rCl}
  \bigHP{\vect{W}^{(m)}}& = &\beta\const{L},\,\forall \,m\in[f],
  \\
  \bigHP{\vect{W}^{(1)},\dots,\vect{W}^{(f)}}& = &f\beta\const{L}\quad (\textnormal{in } q\textnormal{-ary units}).
\end{IEEEeqnarray*}

In PQNMC, a user wishes to privately compute exactly one quadratic nonparallel monomial $X_i^{(k,\ell)}\eqdef W_i^{(k)}W_i^{(\ell)}$, $\forall\,i\in [\beta\const{L}]$, for some $k,\ell\in [f]$, $k < \ell$, out of $\mu\eqdef\binom{f}{2}$ \emph{candidate} quadratic nonparallel monomials. %
For convenience, %
we denote by $\set{T}\eqdef\{(k,\ell)\colon k,\ell\in [f],\,k < \ell\}$ the set of all  ordered $2$-tuples, where $|\set{T}|=\mu$. In $q$-ary units, we have
\begin{IEEEeqnarray*}{rCl}
  \eHP{\vect{X}^{(k,\ell)}}& = &\beta\const{L}\bigHP{X^{(k,\ell)}},\,\forall\,(k,\ell)\in\set{T}, 
  \\
  \bigHP{\vect{X}^{(1,2)},\ldots,\vect{X}^{(f-1,f)}}& = &\beta\const{L}\bigHP{X^{(1,2)},\ldots,X^{(f-1,f)}},
\end{IEEEeqnarray*} 
for \emph{prototype} random variables  $X^{(k,\ell)}$.

The user privately selects an index $(k,\ell)$ and wishes to compute the $(k,\ell)$-th quadratic nonparallel monomial while keeping the requested index $(k,\ell)$ private from each database. In order to retrieve the desired function $\vect{X}^{(k,\ell)}$, $(k,\ell)\in\set{T}$, from the DSS, the user sends a random query to the $j$-th database for all $j\in[n]$. The user generates the queries without any prior knowledge of the realizations of the stored messages, and they are independent of the candidate quadratic nonparallel monomials. In response to the received query, the $j$-th database sends an answer back to the user.

To measure the efficiency of a PC protocol, we consider the required number of downloaded  $q$-ary symbols for retrieving the $\beta\const{L}$  $q$-ary  symbols of the desired function evaluation.
\begin{definition}[PQNMC Rate and Capacity]
  \label{def:def_PCrate}
  The rate of a PQNMC protocol, denoted by $\const{R}$, is defined as the ratio between the \emph{smallest} desired monomial size $\beta\const{L}\HH_\textnormal{min}$ and the total required download cost $\const{D}$, i.e.,
  \begin{IEEEeqnarray*}{c}
    \const{R}\eqdef\frac{\beta\const{L}\HH_\textnormal{min}}{\const{D}},
  \end{IEEEeqnarray*}
  where $\HH_\textnormal{min}\eqdef\min_{(k,\ell)\in\set{T}}{\HP{X^{(k,\ell)}}}$. The PQNMC \emph{capacity}, denoted by $\const{C}_\textnormal{PQNMC}$, is the maximum achievable PQNMC rate over all possible PQNMC protocols.
\end{definition}

Note that for quadratic nonparallel monomials, we have $\HH_\textnormal{max}\eqdef\max_{(k,\ell)\in\set{T}}\HP{X^{(k,\ell)}}=\HH_\textnormal{min}$. Accordingly, every quadratic nonparallel monomial carries the same amount of information and following the terminology of DPIR \cite{ChenWangJafar20_1} we denote the PQNMC problem as \emph{balanced}.

Let $\set{P}(\set{T})$ be the set of all  permutations on the set $\set{T}$, and denote an ordered set $\set{S}\eqdef(\vect{s}_1,\ldots,\vect{s}_\mu)\in\set{P}(\set{T})$. Using the same approach as~\cite{ObeadLinRosnesKliewer19_2}, it can be shown that the PQNMC capacity is bounded from above by $  \const{C}_\textnormal{PQNMC}\leq \overline{\const{C}}(\set{S})$, where
\begin{IEEEeqnarray}{c}
  \overline{\const{C}}(\set{S})\eqdef\frac{n^{\mu}\HH_\textnormal{min}}{\sum\limits_{v=1}^\mu n^{\mu-v+1}\eHPcond{X^{(\vect{s}_v)}}{{X^{(\vect{s}_1)},\dots,X^{(\vect{s}_{v-1})}}}}.\IEEEeqnarraynumspace\label{eq:upper-bound_general-PQNMC-rate}
\end{IEEEeqnarray}
The goal of this work is to determine the best (lowest) outer bound to the PQNMC capacity $\const{C}_\textnormal{PQNMC}$, among all the possible orders of the quadratic nonparallel monomials for a given number of messages $f$, i.e., we are interested in obtaining the best-ordered set that achieves $\min_{\set{S}\in\set{P}(\set{T})}\overline{\const{C}}(\set{S})$.
\begin{remark} 
\label{rem:capacity-outer-bound_independent-n}
We have observed by exhaustive search for $f\leq 5$ that the set of optimal orderings (the ones that minimize  the capacity outer bound in \eqref{eq:upper-bound_general-PQNMC-rate}) is independent of $n \geq 2$, which suggests  that one can choose $n=2$ for finding an optimal order for the capacity outer bound $\overline{\const{C}}(\set{S})$.
\end{remark}

\subsection{Edge-Coloring and Matching}
\label{sec:edge-coloring-and-matching}

Quadratic nonparallel monomials in $f$ variables can be represented by a simple undirected graph $\set{G} = (\set{V},\set{E})$, with $\set{V} = [f]$ where the vertices $k$ and $\ell$ correspond to the messages $W^{(k)}$ and $W^{(\ell)}$ (for \emph{prototype} random variables  $W^{(k)}$ and $W^{(\ell)}$), respectively, and the edge $(k, \ell) \in \set{E}$ represents the nonparallel monomial $X^{(k,\ell)}$. Thus, we have $\mu=|\set{E}|$ and the set of all quadratic nonparallel monomials in $f$ variables are represented by a \emph{complete} graph $\set{K}_f$. 

\begin{definition}[Distance]
Given a graph $\set{G} = (\set{V},\set{E})$, the distance between two vertices $u, v\in \set{V}$, denoted by $d(u,v)$, is the length of the shortest path connecting them, measured in number of edges.  
\end{definition}

\begin{definition}[Matching] 
Given a graph $\set{G} = (\set{V},\set{E})$, a matching $\cal{M}$ in $\set{G}$ is a set of pairwise nonadjacent edges, i.e., a set of edges where no two edges share common vertices. When $|\set{V}|$ is an even number, a \emph{perfect matching} is a matching that includes all vertices of the graph, and when $|\set{V}|$ is an odd number, a \emph{near-perfect matching} is a matching that includes $|\set{V}|-1$ vertices of the graph. 
\end{definition}

\begin{definition}[{Edge-Coloring~\cite[Ch.~17]{BondyMurty08_1}}] %
A \emph{proper} edge-coloring of a graph is an assignment of colors to the edges such that the edges incident to a vertex have distinct colors. 
A graph $\set{G}$ is said to be $\kappa$-edge colorable if $\set{G}$ has a proper edge-coloring with $\kappa$ colors. 
The \emph{chromatic index} of a graph $\set{G}$, denoted by $\chi'(\set{G})$, is the minimum number of colors required to properly edge-color $\set{G}$. 
\end{definition}

The definition of proper edge-coloring of a graph $\set{G}$ implies that the $\kappa$-edge coloring of a graph partitions the graph edge set $\set{E}$ into $\kappa$ (near) perfect matchings $\set{M}_1,\dots,\set{M}_\kappa$ such that $\set{E} = \set{M}_1 \cup\dots\cup\set{M}_\kappa$, and the sets $\set{M}_1,\dots,\set{M}_\kappa$ are known as the color sets of edges.
For complete graphs of $f$ vertices, denoted by $\set{K}_f$, it is known  \cite[Thm.~1]{BehzadChartrandCooper67_1} that %
\ifthenelse{\boolean{short_version}}{
\begin{IEEEeqnarray*}{c}
    \chi^\prime(\set{K}_f)=\begin{cases}
        f-1 & \textnormal{if } f \textnormal{ is even},
        \\
        f   & \textnormal{if } f \textnormal{ is odd}.
    \end{cases}
\end{IEEEeqnarray*}}{
\begin{IEEEeqnarray}{c}
    \chi^\prime(\set{K}_f)=\begin{cases}
        f-1 & \textnormal{if } f \textnormal{ is even},
        \\
        f   & \textnormal{if } f \textnormal{ is odd}.
    \end{cases}
    \label{eq:chromatic-index_complete-graphs}
\end{IEEEeqnarray}}

\begin{remark}
\label{rem:mathcingOrder}
Let $\collect{M}$ be the set of all (near) perfect matchings of a complete graph $\set{K}_f$.
Let $\set{S}[1:\eta]=(\vect{s}_1,\dots, \vect{s}_\eta)$ be the first $\eta$ elements of the ordered set $\set{S}$, where $\eta\eqdef\nicefrac{f(f-1)}{2\chi^\prime(\set{K}_f)}$ is the number of edges within a complete graph matching. For $\set{S}$ to be an \emph{optimal} order of quadratic nonparallel monomials of $f>3$ variables, we conjecture that $\set{S}[1:\eta]$ must constitute a (near) perfect matching of $\set{K}_f$, i.e., $\{\vect{s}_1,\dots, \vect{s}_{\eta}\} \in \collect{M}$. 
\end{remark}

%

\section{Algorithms to Determine a Good Order}
\label{sec:main-results}

In this section, we propose three methods for finding a \emph{good} order, one based on edge-coloring, one based on graph distance, and one entropy-based greedy algorithm.

\begin{remark}
Computing the capacity bound of \eqref{eq:upper-bound_general-PQNMC-rate} for a given order entails the computation of $\mu$ conditional entropies, resulting in a computational complexity of order $\set{O}(q^{\mu})$.
Taking that into account, it can be seen that performing an exhaustive search over all possible orders to optimize the PQNMC capacity outer bound would intuitively require a complexity of order $\set{O}(\mu!\times q^{\mu})$. 
\ifthenelse{\boolean{short_version}}{}{
Nevertheless, due to the balanced nature of the PQNMC candidate function set, i.e., quadratic nonparallel monomials over uniformly distributed random variables, and the resulting symmetry, we would technically require the computation of the capacity outer bound for some $\Delta_f <\mu!$ distinct orders, where
$\Delta_f$ is the total number of distinct paths from the empty graph to the complete graph $\set{K}_f$~\cite[p.~92]{Steinbach04_1}. $\Delta_f$ grows with the number of nonisomorphic graphs with $f$ unlabeled vertices, $\lambda_f$. Finding both $\lambda_f$ and $\Delta_f$ is, in general, a complex problem. However, there exists a collective list of the values of $\lambda_f$ for $f\leq 87$ in~\cite{OEIS-A000088}. For $f=5$, $\lambda_5=34$, and we have numerically computed $\Delta_5=275$, which is significantly smaller than $\mu!=10! = 3628800$.}
\end{remark}

\subsection{Edge-Coloring} %
The key idea is to first find a proper edge-coloring of the complete graph $\set{K}_f$. Then, build an order based on grouping the edges according to their color, i.e., first take the edges corresponding to one of the colors, then the edges corresponding to another color, etc., until all colors have been considered.   
The time complexity of finding an order based on simple edge-coloring follows from the time complexity of finding a proper edge-coloring of $\set{K}_f$. %
We follow the procedure in \ifthenelse{\boolean{short_version}}{\cite[App.~A]{DahliObeadLinRosnes24_1sub}}{Appendix~\ref{append:EdgeColoring}} which runs in polynomial time, i.e., of order ${\mathcal O} (\mu)$.  

We first give an example to illustrate how edge-coloring can give us a good order $\set{S}_\textnormal{EC}$ for the capacity outer bound.

\begin{example}
For $f=6$ messages, a proper $5$-edge coloring of $\set{K}_6$ is as follows:
\label{ex:edge-coloring_f6}
\input{fig/EC_f6.tex}

The resulting order is
\begin{IEEEeqnarray*}{rCl}
   \set{S}_{\textnormal{EC}}& = &\bigl( 
   (1,5), (2,4), (3,6), (1,6), (2,5), (3,4), (1,2),
   \\
   &&\,\, (3,5), (4,6), (1,3), (2,6), (4,5), (1,4), (2,3), (5,6) \bigr),\IEEEeqnarraynumspace
\end{IEEEeqnarray*}
and for $n=2$ and $q=2$, the capacity outer bound is
$\overline{\const{C}}(\set{S}_{\textnormal{EC}}) = 0.5198943946817$. 
\end{example}

The permutation of the colors affects the value of $\overline{\const{C}}$, which indicates that edge-coloring by itself does not guarantee finding an optimal order. Thus, a search over color permutations for the capacity bound, i.e., over $(\chi^\prime(\set{K}_f)-1)!$ permutations (the first $\eta$ edges corresponding to a single color can be fixed; see \Cref{rem:mathcingOrder}), can potentially improve it in exchange for added complexity. We refer to this improved method as enhanced edge-coloring (E-EC), and its complexity is of order $\set{O}(\mu+(\chi^\prime(\set{K}_f)-1)!\times q^{\mu})$. %
For example, if we reorder the colors in Example~\ref{ex:edge-coloring_f6} as (purple, yellow, blue, red, green), we obtain the  order\looseness-1
\begin{IEEEeqnarray*}{rCl}  
   \set{S}_{\textnormal{E-EC}}& = &\bigl((1,5), (2,4), (3,6), (1,6), (2,5), (3,4), 
   (1,4),
   \\
   &&\,(2,3), (5,6), (1,2), (3,5), (4,6), (1,3), (2,6), (4,5)\bigr),\IEEEeqnarraynumspace
\end{IEEEeqnarray*}
which results, for  $n=2$ and $q=2$, in the improved capacity outer bound 
$\overline{\const{C}}(\set{S}_{\textnormal{E-EC}}) = 0.5198121367672$. 

We have observed that even within a set of edges of a given color, the permutation of the edges also affects the value of $\overline{\const{C}}$. For instance, considering the blue color in Example~\ref{ex:edge-coloring_f6}, the values of $\overline{\const{C}}$ between the orders $\{(1,4), (2,3), (5,6)\}$ and $\{(2,3), (1,4), (5,6)\}$ are different. %
However, when searching for the best order within the edges of every color, the computational complexity becomes $\set{O}(\mu+((\eta!)^{\chi^\prime(\set{K}_f)})\times q^{\mu})$, %
which renders finding the best order quickly infeasible. %
Here, we are presenting the (E-)EC solution as a low-complexity solution for finding a \emph{good} order. Thus, we opt out of optimizing the (E-)EC solution any further, and we simply order the edges $(k,\ell)$ within each color set according to the lexicographical order on $[f]\times [f]$. 

The E-EC method is briefly summarized as Algorithm~\ref{alg:best-order_EC-algorithm}.

\begin{algorithm}[t!]
  \caption{Searching for a good order for $\overline{\const{C}}$ based on edge-coloring (E-EC)}
  \label{alg:best-order_EC-algorithm}
  \SetCommentSty{small}
  \DontPrintSemicolon
  \SetKwInOut{Input}{Input}
  \SetKwInOut{Output}{Output}
  \SetKwComment{Comment}{$\triangleright$\ }{}
  \DontPrintSemicolon	 
  
  \Input{$f$, $q$, $n$}
  \Output{A good order of edges $\set{S}_\textnormal{E-EC}$} %
  \small
 $\set{E}_1, \dots \set{E}_{\chi^\prime(\set{K}_f)}\leftarrow$ color sets of edges with edges ordered lexicographically on $[f] \times [f]$\; %
  $\set{E}_\textnormal{E-EC}\leftarrow \set{E}_1$,  $i\leftarrow 1$\;
  $\set{E}_\textnormal{c}\leftarrow$ a permutation of the remaining color sets of edges\;
  $\set{S}_\textnormal{E-EC}\leftarrow (\set{E}_\textnormal{E-EC}\mid\set{E}_\textnormal{c})$\;
  
  Compute $\const{C}_\textnormal{E-EC--best}=\overline{\const{C}}(\set{S}_\textnormal{E-EC})$\;
  \While{$i\leq (\chi^\prime(\set{K}_f)-1)!$}{
    $i\leftarrow i+1$\;
    
    $\set{E}_\textnormal{c}\leftarrow$ next permutation of the remaining color sets of edges\;
    
    \If{$\overline{\const{C}}(\set{E}_\textnormal{E-EC}\mid\set{E}_\textnormal{c})<\const{C}_{\textnormal{E-EC--best}}$}{
      $\set{S}_\textnormal{E-EC}\leftarrow(\set{E}_\textnormal{E-EC}\mid\set{E}_\textnormal{c})$, $\const{C}_\textnormal{E-EC-best} \leftarrow \overline{\const{C}}(\set{S}_\textnormal{E-EC})$\;
    }
  }
  \Return  $\set{S}_\textnormal{E-EC}$
\end{algorithm}
\vspace{-1ex}

\subsection{Longest-Distance First} %
The goal of LDF %
is to minimize dependency among the first selected monomials within an order. Thus, the intuition behind the LDF method also follows from graph matching. However, unlike in the (E-)EC method, we do not restrict ourselves to (near) perfect matchings of the complete graph $\set{K}_f$. 
Here, we follow the convention that if two vertices belong to different connected components, then the distance is defined as infinite~\cite{BondyMurty08_1}, i.e., there is no path connecting the two vertices. The LDF method is summarized with the following sequential steps (further details and a pseudo-code can be found in 
\ifthenelse{\boolean{short_version}}{\cite[App.~B]{DahliObeadLinRosnes24_1sub}}{Appendix~\ref{app:LDF}}).

Start with the null graph $\set{G}=\set{N}_f$, i.e., a graph with $\set{V}=[f]$ and $\set{E}=\emptyset$. Then, adhere to the following steps, adding an edge $(u,v)$ to $\set{G}$ and partial order $\set{S}_{\textnormal{LDF}}$ with each step repetition.
\begin{enumerate}
    \item Repeatedly add an edge not adjacent to any other edge. %
    \item Repeatedly add an edge that connects two vertices with the longest distance and lowest degree in the graph,  until a length-$f$ cycle is formed.
    \item Repeatedly add an edge that connects two vertices with the lexicographically smallest numbers of induced length-$l$ cycles for $3 \leq l \leq f$, until $\set{G}$ is complete, i.e., $\set{G}=\set{K}_f$\ifthenelse{\boolean{short_version}}{.}{ (see \Cref{alg:InnerOrder-algorithm} in Appendix~\ref{app:LDF}).}
\end{enumerate}
As for the (E-)EC  method, we elaborate on the LDF algorithm with an illustrative example. %

\begin{example} 
For $f=6$, %
first, add an edge that is not adjacent to any other edge. For example $(1,2)$, then ${(3,4)}$. As a result, one remaining edge can be added following Step $1)$, which is ${(5,6)}$. Note that $\set{E}=\{(1,2), (3,4), (5,6)\}$ constitute a perfect matching of $\set{K}_6$, adhering to \Cref{rem:mathcingOrder}. 
The corresponding graph $\set{G}$ and graph distances are illustrated in the following figure, where $v\in \set{V}$ denotes the vertex label. 

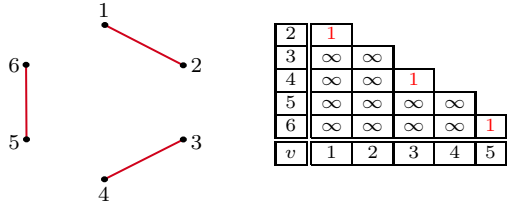
\begin{center}
    \input{fig/LDF_f6_s1_2.tex}
\end{center}

Next, to add an edge that connects two vertices with the longest distance and lowest degree in the graph, we select, for example, the edge $(1,6)$ with $d(1,6)=\infty$ and both vertices of degree $1$. As a result, we have $d(2,5)=3$ and $d(2,6)=d(1,5)=2$, as depicted in the following figure:

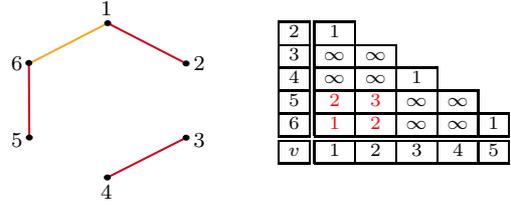
\begin{center}
    \input{fig/LDF_f6_s1_3.tex}
\end{center}

Next, to repeat Step $2)$, we can select from any available edge $(u,v)$ with vertices of degree $1$ and $d(u,v)=\infty$, i.e., $u,v\in \{ 2,3,4,5 \}$. For example, select the edge $(2,3)$, then $(4,5)$ forming a length-$f$ cycle as illustrated in the left-hand side (l.h.s.) of the figure below where a dashed line indicates the order of adding. Now, for Step $3)$, we count the cycles induced from adding each of the remaining edges as seen in the following table, where $\vect{o}=(o_1,\dots, o_i,\dots,o_{f-2})$ and $o_i$ is the number of cycles of length $i+2$. 

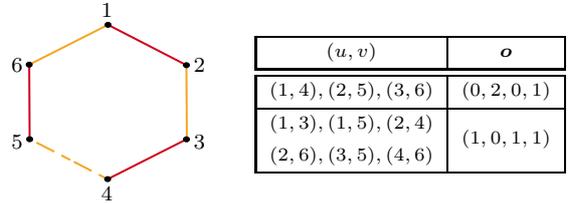
\begin{center}
    \input{fig/LDF_f6_s2_v3.tex}
\end{center}
Each of the edges $(1,4)$, $(2,5)$, and $(3,6)$ induces the smallest number of cycles, lexicographically, thus we select one of these edges. Let that edge be $(1,4)$. The corresponding graph $\set{G}$ is illustrated in the l.h.s of the following figure:\looseness-1
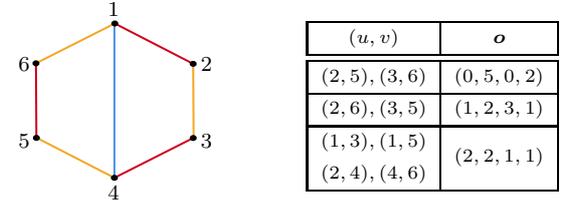
\begin{center}
    \input{fig/LDF_f6_s2_v4.tex}
\end{center}

By repeating Step $3)$, the choices for the following edges remain the same. As can be seen from the above table, edges $(2,5)$ and $(3,6)$ induce the same number of cycles in $\set{G}$. Thus, we select for example $(2,5)$. %
The remaining edges follow from repeating Step $3)$ and are added to the order and $\set{G}$ as illustrated in the following left-to-right top-to-bottom order:  
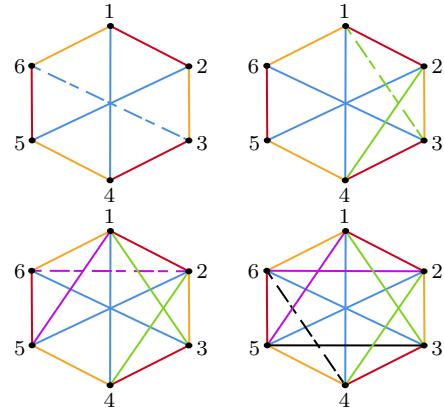
\begin{center}
    \input{fig/LDF_f6_s3.tex}
\end{center}
\begin{center}
    \input{fig/LDF_f6_s4.tex}
\end{center}
\input{izs24PMC_table1}
At the end of the LDF procedure, we obtain the order
\begin{IEEEeqnarray*}{rCl}  
  \set{S}_{\textnormal{LDF}} &=&  \bigl((1,2), (3,4), (5,6), (1,6), (2,3), (4,5), (1,4), \\
     &&\,(2,5), (3,6), (2,4), (1,3), (1,5), (2,6), (3,5), (4,6)\bigr),\IEEEeqnarraynumspace
\end{IEEEeqnarray*}
and for $n=2$ and $q=2$, the capacity outer bound is   $\overline{\const{C}}(\set{S}_{\textnormal{LDF}}) = 0.5197824997350$, which is strictly better than for the E-EC method.
Interestingly,  $\set{S}_{\textnormal{LDF}}$  corresponds to a proper edge-coloring, i.e., no
two adjacent edges share the same color, but it does not correspond to an optimal edge-coloring.
\end{example}

\subsection{Entropy-Based Greedy Method}

The EBG method starts with an empty graph and sequentially adds edges in a greedy manner, i.e., the edge that minimizes the (partial) bound in \eqref{eq:upper-bound_general-PQNMC-rate}, computed based on the new edge and the already added edges, is added at each step in the algorithm. In case of ties, one of the candidate edges is selected at random. In particular, in the first step, an arbitrary edge is added to an initially empty graph. Then, in the second step the bound in \eqref{eq:upper-bound_general-PQNMC-rate} is computed with $\mu=2$ based on the previously added edge and a new edge selected among the possible remaining edges. The new edge that minimizes the computed partial bound is then selected and added to the graph. In this manner, an edge is added to the graph in each step of the algorithm and a monomial order is constructed. The complexity of the EBG method is of order $\set{O}(\nicefrac{\mu(\mu+1)}{2} \times q^{\mu})$.  %
Finally, note that the order returned by the EBG method is independent of the value of $n \in \Naturals$, including $n=1$, while \Cref{{rem:capacity-outer-bound_independent-n}} requires $n \geq 2$. Hence, $n=1$ can be used for better numerical stability when conducting the EBG search.

\section{Discussion and Results}

In \Cref{tab:methodsComparasion}, we compare the results from the proposed (E-)EC, LDF, and EBG methods  for $5 \leq f \leq 12$ messages, for $n=2$ databases, and for a field size of $q=2$ with those of an exhaustive search (for $f=5$) and a directed  random search (for $6 \leq f \leq 12$). The directed  random search is done by first fixing at least the first $f$ edges according to edge-coloring (corresponding to two or three colors) and then conducting a random search among the remaining orders. As can be seen from the table,  the LDF  and  EBG  methods and the exhaustive/directed random search yield the same capacity outer bound for $f \leq 7$ messages, while for $f=8$  and $f=9$  messages a  directed random search gives slightly better results (in the $8$-th digit). (E-)EC gives the same bound as LDF for $f=4$ messages, while for $f \geq 6$, E-EC performs  worse compared to the LDF and EBG methods. The EC method performs in general slightly worse compared to the E-EC method, but has the lowest computational complexity. %
Interestingly, for $f=11$, the LDF method outperforms %
all other methods. The best-known achievable rate $\const{R}$ from \cite[Thm.~2]{ObeadLinRosnesKliewer19_2} is given in the last row of \Cref{tab:methodsComparasion}  to show the cap to the capacity outer bound.
As a final remark, we note  that for larger $q$ (results not included here), the gap between the bounds produced by the LDF and EBG methods increases, which can be attributed to the fact that the proposed simple undirected graph model captures less of the dependencies for larger $q$.\looseness-1

\section{Conclusion}

We proposed two graph-based methods and one EBG algorithm to optimize the order of quadratic monomials in an outer bound for the  PQNMC capacity.  For $f < 6$ messages, all three methods minimize the bound, while for $6 \leq f \leq 12$ the results were compared with those of a directed random search. For almost all  examined cases, the EBG algorithm yields the smallest gap to the best-found  monomial ordering.

\ifthenelse{\boolean{short_version}}{}{
\appendices

\section{Proper Edge-Coloring of a Complete Graph}
\label{append:EdgeColoring}
For completeness, we include in this appendix the procedure followed to properly edge-color a complete graph. 
Consider the complete graph $\set{K}_f$ with vertex and edge sets $\set{V} = [f]$ and $\set{E}=\{(k,\ell): 1\leq k < \ell \leq f\}$, respectively. Recall that $\chi^\prime(\set{K}_f)$ denotes the chromatic index of a complete graph, i.e., the number of colors required to properly edge-color a complete graph, and $\eta$ denotes the number of edges colored with a given color, as given by \eqref{eq:chromatic-index_complete-graphs} and in \Cref{rem:mathcingOrder}, respectively.

\begin{itemize}
    \item  For odd $f$,  let $\set{E}_c$ be the set of edges colored with color $c\in [\chi^\prime(\set{K}_f)]$, where for odd $f$ we have $\chi^\prime(\set{K}_f)=f$. Then, we have $\set{E}_c=\{ (k_p, \ell_p) \,\forall\,  p\in [\eta] \}$ where
      $$k_p, \ell_p \in \{c-p\, \modstar^{*}{f},c+p\,  \modstar^{*}{f} \} $$ such that $k_p<\ell_p$, %
      and $\modstar^{*}{f}$ maps $c-p= 0$ to $f$, but is otherwise equivalent to $\bmod~f$. %
    \item For even $f$, a proper edge-coloring is equivalent to dividing the complete graph $\set{K}_f$ into $f-1$ factors of degree~$1$ \cite[Ch.~XI, Thm.~2]{Koenig90_1}, i.e., $1$-factorization, as follows: the edges $(k, \ell)$ for $k,\ell \in [f-1]$, $k < \ell$, are assigned to color set $\set{E}_c$, for $c\in[\chi^\prime(\set{K}_f)]$, if 
        $$k+\ell-1 \equiv c-1 \pmod{f-1} , $$ 
    and edges $(k,f)$ for $k\in [f-1]$ are assigned to color set $\set{E}_c$, for $c\in[\chi^\prime(\set{K}_f)]$, if 
        $$\; 2k-1 \; \equiv c-1  \pmod{f-1} .$$ 
    \item The elements of the color sets $\set{E}_1, \dots, \set{E}_{\chi^\prime(\set{K}_f)}$ are then ordered lexicographically on $[f] \times [f]$. 
\end{itemize}

\input{izs24PMC_LDF-alg2.tex}

\section{Longest-Distance First Algorithm} \label{app:LDF}
Before we present the LDF algorithm, we remind the reader of some basic graph notation and definitions.

We use the set notation to denote graphs obtained by removing or adding edges from or to a graph $\set{G}=(\set{V},\set{E})$. For example, for  $\set{E}'\subseteq \set{E}$, $\set{G}-\set{E}' \eqdef (\set{V},\set{E}\setminus \set{E}')$,  and, for $\set{E}' \subseteq \{(k,\ell): k,\ell \in \set{V}, k < \ell \}$,  $\set{G}+\set{E}' \eqdef (\set{V},\set{E} \cup \set{E}')$.  Moreover, for a  graph $\set{G}$, we denote by $\set{E}_{\set{G}}$ the set of edges of $\set{G}$ when the underlying graph $\set{G}$ is not unambiguous. The degree of a node $v$ with respect to a graph $\set{G}$ is denoted by $\deg(\set{G},v)$. In the following, the all-zero vector of dimension $a$ is denoted by $\vect{0}_a$ and $[a:b]\eqdef\{a,a+1,\ldots,b\}$ for $a,b\in \Naturals$, $a \leq b$.

\begin{definition}[Eccentricity, Diameter, and Peripherian]
Given a graph $\set{G} = (\set{V},\set{E})$, the eccentricity $\sigma(v)$ of a vertex $v\in \set{V}$ is the maximum distance from $v$ to all other vertices of $\set{G}$, i.e., 
$\sigma(v)= \max_{u\in \set{V}} d(v,u)$.  
The \emph{diameter}  $\diam(\set{G})$ of $\set{G}$ is the maximum eccentricity among all vertices of a graph, i.e., 
$\diam(\set{G})=\max_{v\in \set{V}} {\sigma(v)}. $ 
The \emph{peripherian} of $\set{G}$ is the set of all vertices at the diameter of  $\set{G}$  where each vertex is called peripheral.
\end{definition}

\input{izs24PMC_LDF-alg3.tex}

\Cref{alg:LDF-algorithm} describes the LDF algorithm in details, and it invokes \Cref{alg:InnerOrder-algorithm} and the following sub-routines. 

\begin{itemize}
\item \texttt{Choose-Random-Pair($\set{A}$)}: selects randomly two entires $i,j$ from the input set $\set{A}$ such that $i<j$.
\item \texttt{Connected-Components($\set{G},j)$}: generates a sorted list of connected components of a graph $\set{G}$, i.e., a list of disjoint sets of vertices, in a descending order according to the size of the component and returns the $j$-th set.  
\item \texttt{Periphery($\set{G}$)}: returns a list of vertices at the diameter of $\set{G}$. In line 19 of Algorithm~\ref{alg:LDF-algorithm} this list contains only one pair of vertices. 
\item \texttt{Simple-Cycles($\set{G}$)}: generates a list of simple cycles in a graph $\set{G}$. A simple cycle $\vect{c}$ is a vector of the vertices along a closed path in $\set{G}$ where no vertex appears twice. The length $l$ of a simple cycle $\vect{c}$ is the number of edges in the closed path connecting its vertices, i.e., $|\vect{c}|=l$.
\end{itemize}
Sub-routines that resemble  \texttt{Connected-Components}, \texttt{Periphery}, and \texttt{Simple-Cycles}  can be found in \cite{HagbergSchultSwart08_1}.

Finally, we note that all algorithms targeting cycle enumerating problems can be implemented efficiently following the depth-first search approach.
}

\balance
\bibliographystyle{IEEEtran}
\bibliography{defshort1,biblioHY}

\end{document}

%% file: fig/EC_f6.tex
\tikzset{every picture/.style={line width=0.75pt}} 

\begin{tikzpicture}[x=0.65pt,y=0.6pt,yscale=-1,xscale=1]

\draw [color=darkRed ,draw opacity=1 ]   (278.65,26.55) -- (331.99,56.98) ;
\draw [color=darkYellow  ,draw opacity=1 ]   (278.65,26.55) -- (225.03,56.59) ;
\draw [color=darkYellow ,draw opacity=1 ]   (278.4,143) -- (332.02,112.96) ;
\draw [color=niceGreen  ,draw opacity=1 ]   (278.4,143) -- (225.05,112.57) ;
\draw [color=lightBlue ,draw opacity=1 ]   (225.03,56.59) -- (225.32,112.86) ;
\draw [color=lightBlue  ,draw opacity=1 ]   (331.99,56.98) -- (332.29,112.96) ;
\draw [color=purple  ,draw opacity=1 ]   (225.03,56.59) -- (332.02,112.96) ;
\draw [color=darkYellow  ,draw opacity=1 ]   (331.99,56.98) -- (225.32,112.86) ;
\draw [color=lightBlue  ,draw opacity=1 ]   (278.65,26.55) -- (278.4,143) ;
\draw [color=niceGreen  ,draw opacity=1 ]   (225.03,56.59) -- (331.99,56.98) ;
\draw [color=darkRed  ,draw opacity=1 ]   (225.32,112.86) -- (332.02,112.96) ;
\draw [color=darkRed  ,draw opacity=1 ]   (225.03,56.59) -- (278.4,143) ;
\draw [color=purple  ,draw opacity=1 ]   (331.99,56.98) -- (278.4,143) ;
\draw [color=niceGreen  ,draw opacity=1 ]   (278.65,26.55) -- (332.02,112.96) ;
\draw [color=purple  ,draw opacity=1 ]   (278.65,26.55) -- (225.32,112.86) ;
\draw  [fill=black  ,fill opacity=1 ] (276.96,26.55) .. controls (276.96,25.72) and (277.71,25.05) .. (278.65,25.05) .. controls (279.58,25.05) and (280.34,25.72) .. (280.34,26.55) .. controls (280.34,27.38) and (279.58,28.05) .. (278.65,28.05) .. controls (277.71,28.05) and (276.96,27.38) .. (276.96,26.55) -- cycle ;
\draw  [fill=black  ,fill opacity=1 ] (330.3,56.98) .. controls (330.3,56.15) and (331.06,55.48) .. (331.99,55.48) .. controls (332.93,55.48) and (333.68,56.15) .. (333.68,56.98) .. controls (333.68,57.81) and (332.93,58.48) .. (331.99,58.48) .. controls (331.06,58.48) and (330.3,57.81) .. (330.3,56.98) -- cycle ;
\draw  [fill=black  ,fill opacity=1 ] (330.32,112.96) .. controls (330.32,112.13) and (331.08,111.45) .. (332.02,111.45) .. controls (332.95,111.45) and (333.71,112.13) .. (333.71,112.96) .. controls (333.71,113.79) and (332.95,114.46) .. (332.02,114.46) .. controls (331.08,114.46) and (330.32,113.79) .. (330.32,112.96) -- cycle ;
\draw  [fill=black  ,fill opacity=1 ] (223.34,56.59) .. controls (223.34,55.76) and (224.09,55.09) .. (225.03,55.09) .. controls (225.96,55.09) and (226.72,55.76) .. (226.72,56.59) .. controls (226.72,57.42) and (225.96,58.09) .. (225.03,58.09) .. controls (224.09,58.09) and (223.34,57.42) .. (223.34,56.59) -- cycle ;
\draw  [fill=black  ,fill opacity=1 ] (276.7,143) .. controls (276.7,142.17) and (277.46,141.49) .. (278.4,141.49) .. controls (279.33,141.49) and (280.09,142.17) .. (280.09,143) .. controls (280.09,143.83) and (279.33,144.5) .. (278.4,144.5) .. controls (277.46,144.5) and (276.7,143.83) .. (276.7,143) -- cycle ;
\draw  [fill=black  ,fill opacity=1 ] (223.63,112.86) .. controls (223.63,112.03) and (224.39,111.36) .. (225.32,111.36) .. controls (226.26,111.36) and (227.02,112.03) .. (227.02,112.86) .. controls (227.02,113.69) and (226.26,114.37) .. (225.32,114.37) .. controls (224.39,114.37) and (223.63,113.69) .. (223.63,112.86) -- cycle ;

\draw (15,35) node [anchor=north west][inner sep=0.75pt]    
{\begin{tabular}{c|c}
Purple & $(1,5) (2,4) (3,6)$ \\ 
Yellow & $(1,6) (2,5) (3,4)$ \\
Red & $(1,2) (3,5) (4,6)$ \\ 
Green & $(1,3) (2,6) (4,5)$ \\ 
Blue & $(1,4) (2,3) (5,6)$  
\end{tabular}
};\hfill
\draw (272.96,10.4) node [anchor=north west][inner sep=0.75pt]  [font=\footnotesize]  {$1$};
\draw (337.08,49.71) node [anchor=north west][inner sep=0.75pt]  [font=\footnotesize]  {$2$};
\draw (337.08,107.57) node [anchor=north west][inner sep=0.75pt]  [font=\footnotesize]  {$3$};
\draw (272.96,146.63) node [anchor=north west][inner sep=0.75pt]  [font=\footnotesize]  {$4$};
\draw (210.85,107.57) node [anchor=north west][inner sep=0.75pt]  [font=\footnotesize]  {$5$};
\draw (210,49.71) node [anchor=north west][inner sep=0.75pt]  [font=\footnotesize]  {$6$};

\end{tikzpicture}

%% file: fig/LDF_f6_s1_2.tex

\tikzset{every picture/.style={line width=0.75pt}} 

\begin{tikzpicture}[x=0.55pt,y=0.5pt,yscale=-1,xscale=1]

\draw [color=darkRed, draw opacity=1 ]   (67.65,28.55) -- (120.99,58.98) ;
\draw [color=darkRed, draw opacity=1 ]   (67.4,145) -- (121.02,114.96) ;
\draw [color=darkRed  ,draw opacity=1 ]   (14.03,58.59) -- (14.32,114.86) ;

\draw  [fill=black, fill opacity=1 ] (65.96,28.55) .. controls (65.96,27.72) and (66.71,27.05) .. (67.65,27.05) .. controls (68.58,27.05) and (69.34,27.72) .. (69.34,28.55) .. controls (69.34,29.38) and (68.58,30.05) .. (67.65,30.05) .. controls (66.71,30.05) and (65.96,29.38) .. (65.96,28.55) -- cycle ;
\draw  [fill=black, fill opacity=1 ] (119.3,58.98) .. controls (119.3,58.15) and (120.06,57.48) .. (120.99,57.48) .. controls (121.93,57.48) and (122.68,58.15) .. (122.68,58.98) .. controls (122.68,59.81) and (121.93,60.48) .. (120.99,60.48) .. controls (120.06,60.48) and (119.3,59.81) .. (119.3,58.98) -- cycle ;
\draw  [fill=black, fill opacity=1 ] (119.32,114.96) .. controls (119.32,114.13) and (120.08,113.45) .. (121.02,113.45) .. controls (121.95,113.45) and (122.71,114.13) .. (122.71,114.96) .. controls (122.71,115.79) and (121.95,116.46) .. (121.02,116.46) .. controls (120.08,116.46) and (119.32,115.79) .. (119.32,114.96) -- cycle ;
\draw  [fill=black, fill opacity=1 ] (12.34,58.59) .. controls (12.34,57.76) and (13.09,57.09) .. (14.03,57.09) .. controls (14.96,57.09) and (15.72,57.76) .. (15.72,58.59) .. controls (15.72,59.42) and (14.96,60.09) .. (14.03,60.09) .. controls (13.09,60.09) and (12.34,59.42) .. (12.34,58.59) -- cycle ;
\draw  [fill=black, fill opacity=1 ] (65.7,145) .. controls (65.7,144.17) and (66.46,143.49) .. (67.4,143.49) .. controls (68.33,143.49) and (69.09,144.17) .. (69.09,145) .. controls (69.09,145.83) and (68.33,146.5) .. (67.4,146.5) .. controls (66.46,146.5) and (65.7,145.83) .. (65.7,145) -- cycle ;
\draw  [fill=black, fill opacity=1 ] (12.63,114.86) .. controls (12.63,114.03) and (13.39,113.36) .. (14.32,113.36) .. controls (15.26,113.36) and (16.02,114.03) .. (16.02,114.86) .. controls (16.02,115.69) and (15.26,116.37) .. (14.32,116.37) .. controls (13.39,116.37) and (12.63,115.69) .. (12.63,114.86) -- cycle ;

\draw (61,10) node [anchor=north west][inner sep=0.75pt]  [font=\footnotesize]  {$1$};
\draw (124.08,51.71) node [anchor=north west][inner sep=0.75pt]  [font=\footnotesize]  {$2$};
\draw (124.08,109.57) node [anchor=north west][inner sep=0.75pt]  [font=\footnotesize]  {$3$};
\draw (61,148.63) node [anchor=north west][inner sep=0.75pt]  [font=\footnotesize]  {$4$};
\draw (0,109.57) node [anchor=north west][inner sep=0.75pt]  [font=\footnotesize]  {$5$};
\draw (0,51.71) node [anchor=north west][inner sep=0.75pt]  [font=\footnotesize]  {$6$};

\draw (175,25) node [anchor=north west][inner sep=0.75pt]    
{
\def\arraystretch{1.08}
\setlength{\tabcolsep}{3.5pt}
\scriptsize
\begin{tabular}{|c|@{}c@{\hspace{0.4ex}}|c|c|c|c|c|}
                               \cline{1-1} \cline{3-3}
  $2$ & &{\r$1$}                     \\  \cline{1-1}\cline{3-4}
  $3$ & &$\infty$ & $\infty$                 \\ \cline{1-1}\cline{3-5}
  $4$ & &$\infty$ & $\infty$ & {\r$1$}             \\ \cline{1-1} \cline{3-6}
  $5$ & &$\infty$ & $\infty$ & $\infty$  & $\infty$         \\ \cline{1-1}\cline{3-7}
  $6$ & &$\infty$ & $\infty$ & $\infty$  & $\infty$    & {\r$1$} \\ \cline{1-1} \cline{3-7} \noalign{\vspace{0.5ex}} \cline{1-1}  \cline{3-7}
  $v$ & &$1$      & $2$      & $3$       & $4$         & $5$  \\ \cline{1-1}\cline{3-7}
\end{tabular}
};
\end{tikzpicture}

%% file: fig/LDF_f6_s1_3.tex
\definecolor{darkRed}{RGB}{208,2,27} 
\definecolor{darkYellow}{RGB}{245,166,35}

\tikzset{every picture/.style={line width=0.75pt}} 

\begin{tikzpicture}[x=0.55pt,y=0.5pt,yscale=-1,xscale=1]

\draw [color=darkRed, draw opacity=1 ]   (67.65,28.55) -- (120.99,58.98) ;
\draw [color=darkYellow  ,draw opacity=1 ]  (67.65,28.55) -- (14.32,58.98) ;
\draw [color=darkRed, draw opacity=1 ]   (67.4,145) -- (121.02,114.96) ;
\draw [color=darkRed  ,draw opacity=1 ]   (14.03,58.59) -- (14.32,114.86) ;

\draw  [fill=black, fill opacity=1 ] (65.96,28.55) .. controls (65.96,27.72) and (66.71,27.05) .. (67.65,27.05) .. controls (68.58,27.05) and (69.34,27.72) .. (69.34,28.55) .. controls (69.34,29.38) and (68.58,30.05) .. (67.65,30.05) .. controls (66.71,30.05) and (65.96,29.38) .. (65.96,28.55) -- cycle ;
\draw  [fill=black, fill opacity=1 ] (119.3,58.98) .. controls (119.3,58.15) and (120.06,57.48) .. (120.99,57.48) .. controls (121.93,57.48) and (122.68,58.15) .. (122.68,58.98) .. controls (122.68,59.81) and (121.93,60.48) .. (120.99,60.48) .. controls (120.06,60.48) and (119.3,59.81) .. (119.3,58.98) -- cycle ;
\draw  [fill=black, fill opacity=1 ] (119.32,114.96) .. controls (119.32,114.13) and (120.08,113.45) .. (121.02,113.45) .. controls (121.95,113.45) and (122.71,114.13) .. (122.71,114.96) .. controls (122.71,115.79) and (121.95,116.46) .. (121.02,116.46) .. controls (120.08,116.46) and (119.32,115.79) .. (119.32,114.96) -- cycle ;
\draw  [fill=black, fill opacity=1 ] (12.34,58.59) .. controls (12.34,57.76) and (13.09,57.09) .. (14.03,57.09) .. controls (14.96,57.09) and (15.72,57.76) .. (15.72,58.59) .. controls (15.72,59.42) and (14.96,60.09) .. (14.03,60.09) .. controls (13.09,60.09) and (12.34,59.42) .. (12.34,58.59) -- cycle ;
\draw  [fill=black, fill opacity=1 ] (65.7,145) .. controls (65.7,144.17) and (66.46,143.49) .. (67.4,143.49) .. controls (68.33,143.49) and (69.09,144.17) .. (69.09,145) .. controls (69.09,145.83) and (68.33,146.5) .. (67.4,146.5) .. controls (66.46,146.5) and (65.7,145.83) .. (65.7,145) -- cycle ;
\draw  [fill=black, fill opacity=1 ] (12.63,114.86) .. controls (12.63,114.03) and (13.39,113.36) .. (14.32,113.36) .. controls (15.26,113.36) and (16.02,114.03) .. (16.02,114.86) .. controls (16.02,115.69) and (15.26,116.37) .. (14.32,116.37) .. controls (13.39,116.37) and (12.63,115.69) .. (12.63,114.86) -- cycle ;

\draw (61,10) node [anchor=north west][inner sep=0.75pt]  [font=\footnotesize]  {$1$};
\draw (124.08,51.71) node [anchor=north west][inner sep=0.75pt]  [font=\footnotesize]  {$2$};
\draw (124.08,109.57) node [anchor=north west][inner sep=0.75pt]  [font=\footnotesize]  {$3$};
\draw (61,148.63) node [anchor=north west][inner sep=0.75pt]  [font=\footnotesize]  {$4$};
\draw (0,109.57) node [anchor=north west][inner sep=0.75pt]  [font=\footnotesize]  {$5$};
\draw (0,51.71) node [anchor=north west][inner sep=0.75pt]  [font=\footnotesize]  {$6$};

\draw (175,25) node [anchor=north west][inner sep=0.75pt]    
{
\def\arraystretch{1.08}
\setlength{\tabcolsep}{3.5pt}
\scriptsize
\begin{tabular}{|c|@{}c@{\hspace{0.4ex}}|c|c|c|c|c|}
                                \cline{1-1} \cline{3-3}
  $2$ & & $1$                      \\  \cline{1-1}\cline{3-4}
  $3$ & & $\infty$    & $\infty$                 \\ \cline{1-1} \cline{3-5}
  $4$ & & $\infty$    & $\infty$      & $1$             \\ \cline{1-1} \cline{3-6}
  $5$ & & {\r$2$}    & {\r$3$}       & $\infty$  & $\infty$         \\ \cline{1-1} \cline{3-7}
  $6$ & & {\r$1$}     & {\r$2$}       & $\infty$  & $\infty$      & $1$    \\ \cline{1-1} \cline{3-7} \noalign{\vspace{0.5ex}} \cline{1-1}  \cline{3-7}
  $v$ & & $1$         & $2$           & $3$       & $4$           & $5$  \\ \cline{1-1}\cline{3-7}
\end{tabular}
};
\vspace{-1ex}
\end{tikzpicture}

%% file: fig/LDF_f6_s2_v3.tex

\tikzset{every picture/.style={line width=0.75pt}} 

\begin{tikzpicture}[x=0.55pt,y=0.5pt,yscale=-1,xscale=1]

\draw [color=darkRed  ,draw opacity=1 ]   (67.65,28.55) -- (120.99,58.98) ;
\draw [color=darkYellow, draw opacity=1 ]   (67.65,28.55) -- (14.03,58.59) ;
\draw [color=darkRed  ,draw opacity=1 ] (67.4,145) -- (121.02,114.96) ;
\draw [color=darkYellow, draw opacity=1 ] [dash pattern={on 3.75pt off 3pt on 7.5pt off 1.5pt}]   (67.4,145) -- (14.05,114.57) ;

\draw [color=darkRed ,draw opacity=1 ]   (14.03,58.59) -- (14.32,114.86) ;
\draw [color=darkYellow  ,draw opacity=1 ]   (120.99,58.98) -- (121.29,114.96) ;
\draw  [fill=black  ,fill opacity=1 ] (65.96,28.55) .. controls (65.96,27.72) and (66.71,27.05) .. (67.65,27.05) .. controls (68.58,27.05) and (69.34,27.72) .. (69.34,28.55) .. controls (69.34,29.38) and (68.58,30.05) .. (67.65,30.05) .. controls (66.71,30.05) and (65.96,29.38) .. (65.96,28.55) -- cycle ;
\draw  [fill=black ,fill opacity=1 ] (119.3,58.98) .. controls (119.3,58.15) and (120.06,57.48) .. (120.99,57.48) .. controls (121.93,57.48) and (122.68,58.15) .. (122.68,58.98) .. controls (122.68,59.81) and (121.93,60.48) .. (120.99,60.48) .. controls (120.06,60.48) and (119.3,59.81) .. (119.3,58.98) -- cycle ;
\draw  [fill=black ,fill opacity=1 ] (119.32,114.96) .. controls (119.32,114.13) and (120.08,113.45) .. (121.02,113.45) .. controls (121.95,113.45) and (122.71,114.13) .. (122.71,114.96) .. controls (122.71,115.79) and (121.95,116.46) .. (121.02,116.46) .. controls (120.08,116.46) and (119.32,115.79) .. (119.32,114.96) -- cycle ;
\draw  [fill=black ,fill opacity=1 ] (12.34,58.59) .. controls (12.34,57.76) and (13.09,57.09) .. (14.03,57.09) .. controls (14.96,57.09) and (15.72,57.76) .. (15.72,58.59) .. controls (15.72,59.42) and (14.96,60.09) .. (14.03,60.09) .. controls (13.09,60.09) and (12.34,59.42) .. (12.34,58.59) -- cycle ;
\draw  [fill=black ,fill opacity=1 ] (65.7,145) .. controls (65.7,144.17) and (66.46,143.49) .. (67.4,143.49) .. controls (68.33,143.49) and (69.09,144.17) .. (69.09,145) .. controls (69.09,145.83) and (68.33,146.5) .. (67.4,146.5) .. controls (66.46,146.5) and (65.7,145.83) .. (65.7,145) -- cycle ;
\draw  [fill=black  ,fill opacity=1 ] (12.63,114.86) .. controls (12.63,114.03) and (13.39,113.36) .. (14.32,113.36) .. controls (15.26,113.36) and (16.02,114.03) .. (16.02,114.86) .. controls (16.02,115.69) and (15.26,116.37) .. (14.32,116.37) .. controls (13.39,116.37) and (12.63,115.69) .. (12.63,114.86) -- cycle ;

\draw (61,10) node [anchor=north west][inner sep=0.75pt]  [font=\footnotesize]  {$1$};
\draw (124.08,51.71) node [anchor=north west][inner sep=0.75pt]  [font=\footnotesize]  {$2$};
\draw (124.08,109.57) node [anchor=north west][inner sep=0.75pt]  [font=\footnotesize]  {$3$};
\draw (61,148.63) node [anchor=north west][inner sep=0.75pt]  [font=\footnotesize]  {$4$};
\draw (0,109.57) node [anchor=north west][inner sep=0.75pt]  [font=\footnotesize]  {$5$};
\draw (0,51.71) node [anchor=north west][inner sep=0.75pt]  [font=\footnotesize]  {$6$};

\draw (155,35) node [anchor=north west][inner sep=0.75pt]    
{
\def\arraystretch{1}
\setlength{\tabcolsep}{3.5pt}
\scriptsize
\begin{IEEEeqnarraybox}[
		\IEEEeqnarraystrutmode
		\IEEEeqnarraystrutsizeadd{1.9pt}{2pt}]{v/c/v/c/v}
		\IEEEeqnarrayrulerow\\
		& (u,v) && \vect{o}\\
		\hline\hline
 &(1,4),(2,5),(3,6) && (0, 2, 0, 1)
\\ \hline
&(1,3),(1,5),(2,4) && \multirow{2}{*}{$(1, 0, 1, 1)$} \\ 
&(2,6),(3,5),(4,6) && &
		\\*\IEEEeqnarrayrulerow
\end{IEEEeqnarraybox}

};

\end{tikzpicture}

%% file: fig/LDF_f6_s2_v4.tex

\tikzset{every picture/.style={line width=0.75pt}} 

\begin{tikzpicture}[x=0.55pt,y=0.5pt,yscale=-1,xscale=1]

\draw [color=darkRed  ,draw opacity=1 ]   (67.65,28.55) -- (120.99,58.98) ;
\draw [color=darkYellow, draw opacity=1 ]   (67.65,28.55) -- (14.03,58.59) ;
\draw [color=darkRed  ,draw opacity=1 ] (67.4,145) -- (121.02,114.96) ;
\draw [color=darkYellow, draw opacity=1 ] (67.4,145) -- (14.05,114.57);
\draw [color=darkRed ,draw opacity=1 ]   (14.03,58.59) -- (14.32,114.86) ;
\draw [color=darkYellow  ,draw opacity=1 ]   (120.99,58.98) -- (121.29,114.96) ;
\draw [color=lightBlue, draw opacity=1 ]   (67.4,28.55) -- (67.4,145) ;
\draw  [fill=black  ,fill opacity=1 ] (65.96,28.55) .. controls (65.96,27.72) and (66.71,27.05) .. (67.65,27.05) .. controls (68.58,27.05) and (69.34,27.72) .. (69.34,28.55) .. controls (69.34,29.38) and (68.58,30.05) .. (67.65,30.05) .. controls (66.71,30.05) and (65.96,29.38) .. (65.96,28.55) -- cycle ;
\draw  [fill=black ,fill opacity=1 ] (119.3,58.98) .. controls (119.3,58.15) and (120.06,57.48) .. (120.99,57.48) .. controls (121.93,57.48) and (122.68,58.15) .. (122.68,58.98) .. controls (122.68,59.81) and (121.93,60.48) .. (120.99,60.48) .. controls (120.06,60.48) and (119.3,59.81) .. (119.3,58.98) -- cycle ;
\draw  [fill=black ,fill opacity=1 ] (119.32,114.96) .. controls (119.32,114.13) and (120.08,113.45) .. (121.02,113.45) .. controls (121.95,113.45) and (122.71,114.13) .. (122.71,114.96) .. controls (122.71,115.79) and (121.95,116.46) .. (121.02,116.46) .. controls (120.08,116.46) and (119.32,115.79) .. (119.32,114.96) -- cycle ;
\draw  [fill=black ,fill opacity=1 ] (12.34,58.59) .. controls (12.34,57.76) and (13.09,57.09) .. (14.03,57.09) .. controls (14.96,57.09) and (15.72,57.76) .. (15.72,58.59) .. controls (15.72,59.42) and (14.96,60.09) .. (14.03,60.09) .. controls (13.09,60.09) and (12.34,59.42) .. (12.34,58.59) -- cycle ;
\draw  [fill=black ,fill opacity=1 ] (65.7,145) .. controls (65.7,144.17) and (66.46,143.49) .. (67.4,143.49) .. controls (68.33,143.49) and (69.09,144.17) .. (69.09,145) .. controls (69.09,145.83) and (68.33,146.5) .. (67.4,146.5) .. controls (66.46,146.5) and (65.7,145.83) .. (65.7,145) -- cycle ;
\draw  [fill=black  ,fill opacity=1 ] (12.63,114.86) .. controls (12.63,114.03) and (13.39,113.36) .. (14.32,113.36) .. controls (15.26,113.36) and (16.02,114.03) .. (16.02,114.86) .. controls (16.02,115.69) and (15.26,116.37) .. (14.32,116.37) .. controls (13.39,116.37) and (12.63,115.69) .. (12.63,114.86) -- cycle ;

\draw (61,10) node [anchor=north west][inner sep=0.75pt]  [font=\footnotesize]  {$1$};
\draw (124.08,51.71) node [anchor=north west][inner sep=0.75pt]  [font=\footnotesize]  {$2$};
\draw (124.08,109.57) node [anchor=north west][inner sep=0.75pt]  [font=\footnotesize]  {$3$};
\draw (61,148.63) node [anchor=north west][inner sep=0.75pt]  [font=\footnotesize]  {$4$};
\draw (0,109.57) node [anchor=north west][inner sep=0.75pt]  [font=\footnotesize]  {$5$};
\draw (0,51.71) node [anchor=north west][inner sep=0.75pt]  [font=\footnotesize]  {$6$};

\draw (185,25) node [anchor=north west][inner sep=0.75pt]    
{
\def\arraystretch{1}
\setlength{\tabcolsep}{3.5pt}
\scriptsize
\begin{IEEEeqnarraybox}[
		\IEEEeqnarraystrutmode
		\IEEEeqnarraystrutsizeadd{1.9pt}{2pt}]{v/c/v/c/v}
		\IEEEeqnarrayrulerow\\
		& (u,v) && \vect{o}\\
		\hline\hline
 &(2,5),(3,6) && (0, 5, 0, 2)
\\ \hline
&(2,6),(3,5) && (1, 2, 3, 1)
\\ \hline
&(1,3),(1,5) && \multirow{2}{*}{$(2, 2, 1, 1)$} \\ 
&(2,4),(4,6) && &
		\\*\IEEEeqnarrayrulerow
\end{IEEEeqnarraybox}

};

\end{tikzpicture}

%% file: fig/LDF_f6_s3.tex

\tikzset{every picture/.style={line width=0.75pt}} 

\begin{tikzpicture}[x=0.55pt,y=0.5pt,yscale=-1,xscale=1]

\draw [color=darkRed ,draw opacity=1 ]   (67.65,28.55) -- (120.99,58.98) ;
\draw [color=darkYellow  ,draw opacity=1 ]   (67.65,28.55) -- (14.03,58.59) ;
\draw [color=darkRed ,draw opacity=1 ]   (67.4,145) -- (121.02,114.96) ;
\draw [color=darkYellow  ,draw opacity=1 ]   (67.4,145) -- (14.05,114.57) ;
\draw [color=darkRed ,draw opacity=1 ]   (14.03,58.59) -- (14.32,114.86) ;
\draw [color=darkYellow  ,draw opacity=1 ]   (120.99,58.98) -- (121.29,114.96) ;
\draw [color=lightBlue  ,draw opacity=1 ] [dash pattern={on 3.75pt off 3pt on 7.5pt off 1.5pt}]  (14.03,58.59) -- (121.02,114.96) ;
\draw [color=lightBlue, draw opacity=1 ]   (120.99,58.98) -- (14.32,114.86) ;
\draw [color=lightBlue, draw opacity=1 ]   (67.65,28.55) -- (67.4,145) ;

\draw  [fill=black ,fill opacity=1 ] (65.96,28.55) .. controls (65.96,27.72) and (66.71,27.05) .. (67.65,27.05) .. controls (68.58,27.05) and (69.34,27.72) .. (69.34,28.55) .. controls (69.34,29.38) and (68.58,30.05) .. (67.65,30.05) .. controls (66.71,30.05) and (65.96,29.38) .. (65.96,28.55) -- cycle ;
\draw  [fill=black, fill opacity=1 ] (119.3,58.98) .. controls (119.3,58.15) and (120.06,57.48) .. (120.99,57.48) .. controls (121.93,57.48) and (122.68,58.15) .. (122.68,58.98) .. controls (122.68,59.81) and (121.93,60.48) .. (120.99,60.48) .. controls (120.06,60.48) and (119.3,59.81) .. (119.3,58.98) -- cycle ;
\draw  [fill=black, fill opacity=1 ] (119.32,114.96) .. controls (119.32,114.13) and (120.08,113.45) .. (121.02,113.45) .. controls (121.95,113.45) and (122.71,114.13) .. (122.71,114.96) .. controls (122.71,115.79) and (121.95,116.46) .. (121.02,116.46) .. controls (120.08,116.46) and (119.32,115.79) .. (119.32,114.96) -- cycle ;
\draw  [fill=black,  fill opacity=1 ] (12.34,58.59) .. controls (12.34,57.76) and (13.09,57.09) .. (14.03,57.09) .. controls (14.96,57.09) and (15.72,57.76) .. (15.72,58.59) .. controls (15.72,59.42) and (14.96,60.09) .. (14.03,60.09) .. controls (13.09,60.09) and (12.34,59.42) .. (12.34,58.59) -- cycle ;
\draw  [fill=black, fill opacity=1 ] (65.7,145) .. controls (65.7,144.17) and (66.46,143.49) .. (67.4,143.49) .. controls (68.33,143.49) and (69.09,144.17) .. (69.09,145) .. controls (69.09,145.83) and (68.33,146.5) .. (67.4,146.5) .. controls (66.46,146.5) and (65.7,145.83) .. (65.7,145) -- cycle ;
\draw  [fill={rgb, 255:red, 0; green, 0; blue, 0 }  ,fill opacity=1 ] (12.63,114.86) .. controls (12.63,114.03) and (13.39,113.36) .. (14.32,113.36) .. controls (15.26,113.36) and (16.02,114.03) .. (16.02,114.86) .. controls (16.02,115.69) and (15.26,116.37) .. (14.32,116.37) .. controls (13.39,116.37) and (12.63,115.69) .. (12.63,114.86) -- cycle ;
\draw [color={rgb, 255:red, 208; green, 2; blue, 27 }  ,draw opacity=1 ]   (227.65,28.55) -- (280.99,58.98) ;
\draw [color=darkYellow  ,draw opacity=1 ]   (227.65,28.55) -- (174.03,58.59) ;
\draw [color={rgb, 255:red, 208; green, 2; blue, 27 }  ,draw opacity=1 ]   (227.4,145) -- (281.02,114.96) ;
\draw [color=darkYellow  ,draw opacity=1 ]   (227.4,145) -- (174.05,114.57) ;
\draw [color={rgb, 255:red, 208; green, 2; blue, 27 }  ,draw opacity=1 ]   (174.03,58.59) -- (174.32,114.86) ;
\draw [color=darkYellow  ,draw opacity=1 ]   (280.99,58.98) -- (281.29,114.96) ;
\draw [color=lightBlue, draw opacity=1 ]   (174.03,58.59) -- (281.02,114.96) ;
\draw [color=lightBlue, draw opacity=1 ]   (280.99,58.98) -- (174.32,114.86) ;
\draw [color=lightBlue, draw opacity=1 ]   (227.65,28.55) -- (227.4,145) ;
\draw [color=niceGreen ,draw opacity=1 ]   (280.99,58.98) -- (227.4,145) ;
\draw [color=niceGreen ,draw opacity=1 ] [dash pattern={on 3.75pt off 3pt on 7.5pt off 1.5pt}]  (227.65,28.55) -- (281.02,114.96) ;

\draw  [fill=black, fill opacity=1 ] (225.96,28.55) .. controls (225.96,27.72) and (226.71,27.05) .. (227.65,27.05) .. controls (228.58,27.05) and (229.34,27.72) .. (229.34,28.55) .. controls (229.34,29.38) and (228.58,30.05) .. (227.65,30.05) .. controls (226.71,30.05) and (225.96,29.38) .. (225.96,28.55) -- cycle ;
\draw  [fill=black, fill opacity=1 ] (279.3,58.98) .. controls (279.3,58.15) and (280.06,57.48) .. (280.99,57.48) .. controls (281.93,57.48) and (282.68,58.15) .. (282.68,58.98) .. controls (282.68,59.81) and (281.93,60.48) .. (280.99,60.48) .. controls (280.06,60.48) and (279.3,59.81) .. (279.3,58.98) -- cycle ;
\draw  [fill=black, fill opacity=1 ] (279.32,114.96) .. controls (279.32,114.13) and (280.08,113.45) .. (281.02,113.45) .. controls (281.95,113.45) and (282.71,114.13) .. (282.71,114.96) .. controls (282.71,115.79) and (281.95,116.46) .. (281.02,116.46) .. controls (280.08,116.46) and (279.32,115.79) .. (279.32,114.96) -- cycle ;
\draw  [fill=black, fill opacity=1 ] (172.34,58.59) .. controls (172.34,57.76) and (173.09,57.09) .. (174.03,57.09) .. controls (174.96,57.09) and (175.72,57.76) .. (175.72,58.59) .. controls (175.72,59.42) and (174.96,60.09) .. (174.03,60.09) .. controls (173.09,60.09) and (172.34,59.42) .. (172.34,58.59) -- cycle ;
\draw  [fill=black, fill opacity=1 ] (225.7,145) .. controls (225.7,144.17) and (226.46,143.49) .. (227.4,143.49) .. controls (228.33,143.49) and (229.09,144.17) .. (229.09,145) .. controls (229.09,145.83) and (228.33,146.5) .. (227.4,146.5) .. controls (226.46,146.5) and (225.7,145.83) .. (225.7,145) -- cycle ;
\draw  [fill=black, fill opacity=1 ] (172.63,114.86) .. controls (172.63,114.03) and (173.39,113.36) .. (174.32,113.36) .. controls (175.26,113.36) and (176.02,114.03) .. (176.02,114.86) .. controls (176.02,115.69) and (175.26,116.37) .. (174.32,116.37) .. controls (173.39,116.37) and (172.63,115.69) .. (172.63,114.86) -- cycle ;

\draw (61,10) node [anchor=north west][inner sep=0.75pt]  [font=\footnotesize]  {$1$};
\draw (124.08,51.71) node [anchor=north west][inner sep=0.75pt]  [font=\footnotesize]  {$2$};
\draw (124.08,109.57) node [anchor=north west][inner sep=0.75pt]  [font=\footnotesize]  {$3$};
\draw (61,148.63) node [anchor=north west][inner sep=0.75pt]  [font=\footnotesize]  {$4$};
\draw (0,109.57) node [anchor=north west][inner sep=0.75pt]  [font=\footnotesize]  {$5$};
\draw (0,51.71) node [anchor=north west][inner sep=0.75pt]  [font=\footnotesize]  {$6$};
\draw (221,10) node [anchor=north west][inner sep=0.75pt]  [font=\footnotesize]  {$1$};
\draw (284.08,51.71) node [anchor=north west][inner sep=0.75pt]  [font=\footnotesize]  {$2$};
\draw (284.08,109.57) node [anchor=north west][inner sep=0.75pt]  [font=\footnotesize]  {$3$};
\draw (221,148.63) node [anchor=north west][inner sep=0.75pt]  [font=\footnotesize]  {$4$};
\draw (159.85,109.57) node [anchor=north west][inner sep=0.75pt]  [font=\footnotesize]  {$5$};
\draw (159,51.71) node [anchor=north west][inner sep=0.75pt]  [font=\footnotesize]  {$6$};

\end{tikzpicture}

%% file: fig/LDF_f6_s4.tex
\tikzset{every picture/.style={line width=0.75pt}} 

\begin{tikzpicture}[x=0.55pt,y=0.5pt,yscale=-1,xscale=1]

\draw [color=darkRed ,draw opacity=1 ]   (67.65,28.55) -- (120.99,58.98) ;
\draw [color=darkYellow ,draw opacity=1 ]   (67.65,28.55) -- (14.03,58.59) ;
\draw [color=darkRed ,draw opacity=1 ]   (67.4,145) -- (121.02,114.96) ;
\draw [color=darkYellow ,draw opacity=1 ]   (67.4,145) -- (14.05,114.57) ;
\draw [color=darkRed ,draw opacity=1 ]   (14.03,58.59) -- (14.32,114.86) ;
\draw [color=darkYellow ,draw opacity=1 ]   (120.99,58.98) -- (121.29,114.96) ;
\draw [color=lightBlue  ,draw opacity=1 ]   (14.03,58.59) -- (121.02,114.96) ;
\draw [color=lightBlue  ,draw opacity=1 ]   (120.99,58.98) -- (14.32,114.86) ;
\draw [color=lightBlue  ,draw opacity=1 ]   (67.65,28.55) -- (67.4,145) ;
\draw [color=purple  ,draw opacity=1 ] [dash pattern={on 3.75pt off 3pt on 7.5pt off 1.5pt}]  (14.03,58.59) -- (120.99,58.98) ;
\draw [color=niceGreen ,draw opacity=1 ]   (120.99,58.98) -- (67.4,145) ;
\draw [color=niceGreen ,draw opacity=1 ]   (67.65,28.55) -- (121.02,114.96) ;
\draw [color=purple  ,draw opacity=1 ]   (67.65,28.55) -- (14.32,114.86) ;
\draw  [fill=black  ,fill opacity=1 ] (65.96,28.55) .. controls (65.96,27.72) and (66.71,27.05) .. (67.65,27.05) .. controls (68.58,27.05) and (69.34,27.72) .. (69.34,28.55) .. controls (69.34,29.38) and (68.58,30.05) .. (67.65,30.05) .. controls (66.71,30.05) and (65.96,29.38) .. (65.96,28.55) -- cycle ;
\draw  [fill=black  ,fill opacity=1 ] (119.3,58.98) .. controls (119.3,58.15) and (120.06,57.48) .. (120.99,57.48) .. controls (121.93,57.48) and (122.68,58.15) .. (122.68,58.98) .. controls (122.68,59.81) and (121.93,60.48) .. (120.99,60.48) .. controls (120.06,60.48) and (119.3,59.81) .. (119.3,58.98) -- cycle ;
\draw  [fill=black  ,fill opacity=1 ] (119.32,114.96) .. controls (119.32,114.13) and (120.08,113.45) .. (121.02,113.45) .. controls (121.95,113.45) and (122.71,114.13) .. (122.71,114.96) .. controls (122.71,115.79) and (121.95,116.46) .. (121.02,116.46) .. controls (120.08,116.46) and (119.32,115.79) .. (119.32,114.96) -- cycle ;
\draw  [fill=black  ,fill opacity=1 ] (12.34,58.59) .. controls (12.34,57.76) and (13.09,57.09) .. (14.03,57.09) .. controls (14.96,57.09) and (15.72,57.76) .. (15.72,58.59) .. controls (15.72,59.42) and (14.96,60.09) .. (14.03,60.09) .. controls (13.09,60.09) and (12.34,59.42) .. (12.34,58.59) -- cycle ;
\draw  [fill=black  ,fill opacity=1 ] (65.7,145) .. controls (65.7,144.17) and (66.46,143.49) .. (67.4,143.49) .. controls (68.33,143.49) and (69.09,144.17) .. (69.09,145) .. controls (69.09,145.83) and (68.33,146.5) .. (67.4,146.5) .. controls (66.46,146.5) and (65.7,145.83) .. (65.7,145) -- cycle ;
\draw  [fill=black  ,fill opacity=1 ] (12.63,114.86) .. controls (12.63,114.03) and (13.39,113.36) .. (14.32,113.36) .. controls (15.26,113.36) and (16.02,114.03) .. (16.02,114.86) .. controls (16.02,115.69) and (15.26,116.37) .. (14.32,116.37) .. controls (13.39,116.37) and (12.63,115.69) .. (12.63,114.86) -- cycle ;

\draw [color=darkRed ,draw opacity=1 ]   (227.65,28.55) -- (280.99,58.98) ;
\draw [color=darkYellow ,draw opacity=1 ]   (227.65,28.55) -- (174.03,58.59) ;
\draw [color=darkRed ,draw opacity=1 ]   (227.4,145) -- (281.02,114.96) ;
\draw [color=darkYellow ,draw opacity=1 ]   (227.4,145) -- (174.05,114.57) ;
\draw [color=darkRed ,draw opacity=1 ]   (174.03,58.59) -- (174.32,114.86) ;
\draw [color=darkYellow ,draw opacity=1 ]   (280.99,58.98) -- (281.29,114.96) ;
\draw [color=lightBlue  ,draw opacity=1 ]   (174.03,58.59) -- (281.02,114.96) ;
\draw [color=lightBlue  ,draw opacity=1 ]   (280.99,58.98) -- (174.32,114.86) ;
\draw [color=lightBlue  ,draw opacity=1 ]   (227.65,28.55) -- (227.4,145) ;
\draw [color=purple  ,draw opacity=1 ]   (174.03,58.59) -- (280.99,58.98) ;
\draw    (174.32,114.86) -- (281.02,114.96) ;
\draw  [dash pattern={on 3.75pt off 3pt on 7.5pt off 1.5pt}]  (174.03,58.59) -- (227.4,145) ;
\draw [color=niceGreen ,draw opacity=1 ]   (280.99,58.98) -- (227.4,145) ;
\draw [color=niceGreen ,draw opacity=1 ]   (227.65,28.55) -- (281.02,114.96) ;
\draw [color=purple  ,draw opacity=1 ]   (227.65,28.55) -- (174.32,114.86) ;
\draw  [fill=black  ,fill opacity=1 ] (225.96,28.55) .. controls (225.96,27.72) and (226.71,27.05) .. (227.65,27.05) .. controls (228.58,27.05) and (229.34,27.72) .. (229.34,28.55) .. controls (229.34,29.38) and (228.58,30.05) .. (227.65,30.05) .. controls (226.71,30.05) and (225.96,29.38) .. (225.96,28.55) -- cycle ;
\draw  [fill=black  ,fill opacity=1 ] (279.3,58.98) .. controls (279.3,58.15) and (280.06,57.48) .. (280.99,57.48) .. controls (281.93,57.48) and (282.68,58.15) .. (282.68,58.98) .. controls (282.68,59.81) and (281.93,60.48) .. (280.99,60.48) .. controls (280.06,60.48) and (279.3,59.81) .. (279.3,58.98) -- cycle ;
\draw  [fill=black  ,fill opacity=1 ] (279.32,114.96) .. controls (279.32,114.13) and (280.08,113.45) .. (281.02,113.45) .. controls (281.95,113.45) and (282.71,114.13) .. (282.71,114.96) .. controls (282.71,115.79) and (281.95,116.46) .. (281.02,116.46) .. controls (280.08,116.46) and (279.32,115.79) .. (279.32,114.96) -- cycle ;
\draw  [fill=black  ,fill opacity=1 ] (172.34,58.59) .. controls (172.34,57.76) and (173.09,57.09) .. (174.03,57.09) .. controls (174.96,57.09) and (175.72,57.76) .. (175.72,58.59) .. controls (175.72,59.42) and (174.96,60.09) .. (174.03,60.09) .. controls (173.09,60.09) and (172.34,59.42) .. (172.34,58.59) -- cycle ;
\draw  [fill=black  ,fill opacity=1 ] (225.7,145) .. controls (225.7,144.17) and (226.46,143.49) .. (227.4,143.49) .. controls (228.33,143.49) and (229.09,144.17) .. (229.09,145) .. controls (229.09,145.83) and (228.33,146.5) .. (227.4,146.5) .. controls (226.46,146.5) and (225.7,145.83) .. (225.7,145) -- cycle ;
\draw  [fill=black  ,fill opacity=1 ] (172.63,114.86) .. controls (172.63,114.03) and (173.39,113.36) .. (174.32,113.36) .. controls (175.26,113.36) and (176.02,114.03) .. (176.02,114.86) .. controls (176.02,115.69) and (175.26,116.37) .. (174.32,116.37) .. controls (173.39,116.37) and (172.63,115.69) .. (172.63,114.86) -- cycle ;

\draw (61,10) node [anchor=north west][inner sep=0.75pt]  [font=\footnotesize]  {$1$};
\draw (124.08,51.71) node [anchor=north west][inner sep=0.75pt]  [font=\footnotesize]  {$2$};
\draw (124.08,109.57) node [anchor=north west][inner sep=0.75pt]  [font=\footnotesize]  {$3$};
\draw (61,148.63) node [anchor=north west][inner sep=0.75pt]  [font=\footnotesize]  {$4$};
\draw (0,109.57) node [anchor=north west][inner sep=0.75pt]  [font=\footnotesize]  {$5$};
\draw (0,51.71) node [anchor=north west][inner sep=0.75pt]  [font=\footnotesize]  {$6$};
\draw (221,10) node [anchor=north west][inner sep=0.75pt]  [font=\footnotesize]  {$1$};
\draw (284.08,51.71) node [anchor=north west][inner sep=0.75pt]  [font=\footnotesize]  {$2$};
\draw (284.08,109.57) node [anchor=north west][inner sep=0.75pt]  [font=\footnotesize]  {$3$};
\draw (221,148.63) node [anchor=north west][inner sep=0.75pt]  [font=\footnotesize]  {$4$};
\draw (159.85,109.57) node [anchor=north west][inner sep=0.75pt]  [font=\footnotesize]  {$5$};
\draw (159,51.71) node [anchor=north west][inner sep=0.75pt]  [font=\footnotesize]  {$6$};

\end{tikzpicture}

%% file: izs24PMC_table1.tex
\begin{table*}[t]
	\centering
	\caption{Comparison between PQNMC capacity outer bounds obtained with the (E-)EC ($\overline{\const{C}}(\set{S}_{\textnormal{(E-)EC}})$),  LDF ($\overline{\const{C}}(\set{S}_{\textnormal{LDF}})$), and EBG ($\overline{\const{C}}(\set{S}_{\textnormal{EBG}})$) methods, as well as with the best bound found by exhaustive/directed random search ($\overline{\const{C}}(\set{S}_{\textnormal{ES/RS}})$), for $n=2$ databases, and for a field size of $q=2$. The best bound for each number of messages $f$ is marked in bold. The best-known achievable rate $\const{R}$ from \cite[Thm.~2]{ObeadLinRosnesKliewer19_2} is given as well to show the cap to the capacity outer bound.}
	\label{tab:methodsComparasion}
    \vskip -2mm
    \Resize[2.03\columnwidth]{
	\begin{IEEEeqnarraybox*}[
		\IEEEeqnarraystrutmode
		\IEEEeqnarraystrutsizeadd{3pt}{2pt}]{c/c/c/c/c/c/c/c/c} 
		\toprule
		f 
		& 5 & 6 & 7 & 8 & 9 & 10& 11 & 12\\
		\midrule 
		\overline{\const{C}}(\set{S}_{\textnormal{EC}})
		& 0.5382035621102 
		& 0.5198943946817 
		& 0.5158988408975 
		& 0.5088200966114 
		& 0.5071434701312 
		& 0.5041602427037 
		& 0.5033789063480 
		& 0.5020207578041 
		\\
		\overline{\const{C}}(\set{S}_{\textnormal{E-EC}})\setfootnotemark\label{first}
		& \bf{0.5321513151313}  
		& 0.5198121367672 
		& 0.5130098344723 
		& 0.5085684044374 
		& 0.5058885273733 
		& 0.5039972538181 
		& 0.5028028499055 
		& 0.5019311781396 
		\\   
		\overline{\const{C}}(\set{S}_{\textnormal{LDF}})
		& \bf{0.5321513151313}  
		& \bf{0.5197824997350}  
		& \bf{0.5129571653366} 
		& 0.5085546467521 
		& 0.5058724664437 
		& 0.5039960955809 
		& \bf{0.5027529132784} 
		& 0.5019069907637 
		\\ 
		\overline{\const{C}}(\set{S}_{\textnormal{EBG}})
		& \bf{0.5321513151313} 
		& \bf{0.5197824997350} 
		& \bf{0.5129571653366} 
		& 0.5085546463038  
		& 0.5058724626997 
		& \bf{0.5039958945996} 
		& 0.5027582097217 
		& \bf{0.5019068074415} 
		\\
		\overline{\const{C}}(\set{S}_{\textnormal{ES/RS}}) 
		& \bf{0.5321513151313} 
		& \bf{0.5197824997350} 
		& \bf{0.5129571653366} 
		& \bf{0.5085546398430} 
		& \bf{0.5058724411573} 
		& 0.5039961304091 
		& 0.5027529200313 
		& 0.5019070293099 
		\\ 
        \const{R}
		& 0.5026676304668  
		& 0.5001371033940  
		& 0.5000032431709  
		& 0.5000000359051  
		& 0.5000000001891  
		& 0.5000000000005  
		& 0.5000000000000  
		& 0.5000000000000  
		\\   
		\bottomrule
\end{IEEEeqnarraybox*}
}

\afterpage{\footnotetext[\getrefnumber{first}]{Due to high computational complexity, we fix two color sets of edges and then search over the remaining color permutations for $f=11$ and $f=12$.
\vspace{-3ex}} }
\end{table*}

%% file: izs24PMC_LDF-alg2.tex
\begin{algorithm}[t!] %
	\caption{Searching for a good order for $\overline{\const{C}}$ based on graph eccentricity and induced cycles (LDF)}
	\label{alg:LDF-algorithm}
	\SetCommentSty{small}
	\DontPrintSemicolon
	\SetKwFunction{AddEdge}{Add-Edge}
	\SetKwFunction{RemoveEdge}{Remove-Edge}
	\SetKwFunction{ChooseRandomP}{Choose-Random-Pair}
	\SetKwFunction{CC}{Connected-Components}
	\SetKwFunction{Periphery}{Periphery}
	\SetKwFunction{OrderInnerEdges}{Order-Inner-Edges}
	\SetKwInOut{Input}{Input}
	\SetKwInOut{Output}{Output}
	\SetKwComment{Comment}{$\triangleright$\ }{}{}
	\DontPrintSemicolon	 
	\Input{$f$}
	\Output{$\set{S}_\textnormal{LDF}$ }
	\small
	\Comment{\b start with a random edge or, alternatively, fix the starting edge to $(1,2)$}
	$(u,v)\leftarrow \ChooseRandomP([f]) $\; 
	$\set{G}\leftarrow \set{N}_f+ \{(u, v)\}$,
	$\set{S}_\textnormal{OE} \leftarrow \bigl((u,v)\bigr) $\;
	\Comment{\b order of the outer edges of a complete graph, up to an isomorphism}
	\Comment{\b add edges that create a connected graph with maximum graph eccentricity}
	\For{$1 \leq i \leq  f-2$}{
		\For{$1 \leq j\leq 2$} {
			$\set{V}_j \leftarrow \CC(\set{G},j)$\;
			$min_d \leftarrow 2$\;
			\For{$v\in \set{V}_j$}{
				\If{$\deg(\set{G},v)< min_d$}{
					$min_d \leftarrow \deg(\set{G},v)$\;
					$v_j \leftarrow v$\;}
			}
		}
		$\set{G}\leftarrow \set{G}+ \{(v_1, v_2)\}$,
		$\set{S}_\textnormal{OE}\leftarrow (\set{S}_\textnormal{OE}\mid (v_1,v_2))$\;
	}
	$(v_1,v_2)\leftarrow \Periphery(\set{G})$\;
	$\set{G}\leftarrow \set{G}+\{(v_1, v_2)\}$,
	$\set{S}_\textnormal{OE}\leftarrow (\set{S}_\textnormal{OE}\mid (v_1,v_2))$\;
	$\set{S}_\textnormal{LDF}, \set{G} \leftarrow \OrderInnerEdges(f, \set{S}_\textnormal{OE},\set{G})$\;
	\Return $\set{S}_\textnormal{LDF}$ 
\end{algorithm}

%% file: izs24PMC_LDF-alg3.tex
\begin{algorithm}[t!] %
	\caption{\texttt{Order-Inner-Edges}}
	\label{alg:InnerOrder-algorithm}
	\SetCommentSty{small}
	\DontPrintSemicolon
	\SetKwFunction{RemoveEdge}{Remove-Edge}
	\SetKwFunction{AddEdge}{Add-Edge}
	\SetKwFunction{SimpleCycles}{Simple-Cycles}
	\SetKwFunction{OrderInnerEdges}{Order-Inner-Edges}
	\SetKwInOut{Input}{Input}
	\SetKwInOut{Output}{Output}
	\SetKwComment{Comment}{$\triangleright$\ }{}{}
	\DontPrintSemicolon	 
	\Input{$f, \set{S}_\textnormal{OE}, \set{G} $}
	\Output{An updated graph $\set{G}$ and a partial order $\set{S}_\textnormal{PO}$ after adding an edge from the inner edges of $\set{K}_f$ that minimizes lexicographically the numbers of induced length-$l$ cycles in $\set{G}$}
	\small
	$\set{E}_\textnormal{d} \leftarrow \set{E}_{\set{K}_f}\setminus \set{E}_{\set{G}}$ \Comment{\b set of  candidate edges}
	\Comment{\b count the number of cycles induced by adding an edge for every edge in $\set{E}_\textnormal{d}$}
	\For{$ (v_1,v_2) \in  \set{E}_\textnormal{d} $}{
		\Comment{\b create a vector of counters for cycles of all lengths}
		$\vect{o}^{(v_1,v_2)}=(o_1, \dots, o_{f-2}) \leftarrow \vect{0}_{f-2} $\;
		$\set{G}\leftarrow \set{G}+\{(v_1, v_2)\}$ \;
		$ \collect{G}_{\textnormal{cyc}} \leftarrow \SimpleCycles(\set{G})$\;  
		\Comment{\b count number of cycles of different lengths $l \in [3:f]$}
		\For{$\vect{c} \in \collect{G}_{\textnormal{cyc}}$}{
			$l \leftarrow|\vect{c}|$\;
			$o_{l-2}\leftarrow o_{l-2}+1 $ 
		}
		$\set{G}\leftarrow \set{G}-\{(v_1, v_2)\}$
	}
	\Comment{\b find and add the edge with minimum number of smaller cycles by comparing the vectors $\vect{o}^{(v_1,v_2)}$ lexicographically} 
	$(\hat{v}_1, \hat{v}_2) \leftarrow \argmin_{(v_1,v_2)} \{ \vect{o}^{(v_1,v_2)} \} $\;
	$\set{G}\leftarrow \set{G}+\{(\hat{v}_1, \hat{v}_2)\}$,
	$\set{S}_\textnormal{PO}\leftarrow (\set{S}_\textnormal{OE} \mid (\hat{v}_1, \hat{v}_2))$\;
	\Comment{\b iterate over remaining edges}
	\uIf{$|\set{E}_\textnormal{d} | \neq 2$}{
		$\set{S}_\textnormal{PO}, \set{G}\leftarrow \OrderInnerEdges(f,\set{S}_\textnormal{PO}, \set{G})$
	}\Else 
	{ \Comment{\b add the last two edges in any order}
		\For{$(v_1,v_2) \in  \set{E}_\textnormal{d} $}
		{
			$\set{G}\leftarrow \set{G}+\{(v_1, v_2)\}$, 
			$\set{S}_\textnormal{PO}\leftarrow (\set{S}_\textnormal{PO}\mid (v_1, v_2))$\;
		}
	} 
	\Return $\set{S}_\textnormal{PO}, \set{G}$
\end{algorithm}

%% file: izs24PMC_v4.bbl
\begin{thebibliography}{10}
\providecommand{\url}[1]{#1}
\csname url@samestyle\endcsname
\providecommand{\newblock}{\relax}
\providecommand{\bibinfo}[2]{#2}
\providecommand{\BIBentrySTDinterwordspacing}{\spaceskip=0pt\relax}
\providecommand{\BIBentryALTinterwordstretchfactor}{4}
\providecommand{\BIBentryALTinterwordspacing}{\spaceskip=\fontdimen2\font plus
\BIBentryALTinterwordstretchfactor\fontdimen3\font minus
  \fontdimen4\font\relax}
\providecommand{\BIBforeignlanguage}[2]{{%
\expandafter\ifx\csname l@#1\endcsname\relax
\typeout{** WARNING: IEEEtran.bst: No hyphenation pattern has been}%
\typeout{** loaded for the language `#1'. Using the pattern for}%
\typeout{** the default language instead.}%
\else
\language=\csname l@#1\endcsname
\fi
#2}}
\providecommand{\BIBdecl}{\relax}
\BIBdecl

\bibitem{SunJafar19_2}
H.~Sun and S.~A. Jafar, ``The capacity of private computation,'' \emph{IEEE
  Trans. Inf. Theory}, vol.~65, no.~6, pp. 3880--3897, Jun. 2019.

\bibitem{ObeadKliewer18_1}
S.~A. Obead and J.~Kliewer, ``Achievable rate of private function retrieval
  from {MDS} coded databases,'' in \emph{Proc. IEEE Int. Symp. Inf. Theory
  (ISIT)}, Vail, CO, USA, Jun. 17--22, 2018, pp. 2117--2121.

\bibitem{ObeadLinRosnesKliewer22_1}
S.~A. Obead, H.-Y. Lin, E.~Rosnes, and J.~Kliewer, ``Private linear computation
  for noncolluding coded databases,'' \emph{IEEE J. Sel. Areas Commun.},
  vol.~40, no.~3, pp. 847--861, Mar. 2022.

\bibitem{Karpuk18_1}
D.~Karpuk, ``Private computation of systematically encoded data with colluding
  servers,'' in \emph{Proc. IEEE Int. Symp. Inf. Theory (ISIT)}, Vail, CO, USA,
  Jun. 17--22, 2018, pp. 2112--2116.

\bibitem{RavivKarpuk19_2}
N.~Raviv and D.~A. Karpuk, ``Private polynomial computation from {L}agrange
  encoding,'' \emph{IEEE Trans. Inf. Forens. Secur.}, vol.~15, pp. 553--563,
  2020.

\bibitem{ObeadLinRosnesKliewer22_2}
S.~A. Obead, H.-Y. Lin, E.~Rosnes, and J.~Kliewer, ``Private polynomial
  function computation for noncolluding coded databases,'' \emph{IEEE Trans.
  Inf. Forens. Secur.}, vol.~17, pp. 1800--1813, 2022.

\bibitem{YakimenkaLinRosnes20_1}
Y.~Yakimenka, H.-Y. Lin, and E.~Rosnes, ``On the capacity of private monomial
  computation,'' in \emph{Proc. Int. Zurich Sem. Inf. Commun. (IZS)}, Zurich,
  Switzerland, Feb. 26--28, 2020, pp. 31--35.

\bibitem{ZhuYanTangLi22_1}
J.~Zhu, Q.~Yan, X.~Tang, and S.~Li, ``Symmetric private polynomial computation
  from {Lagrange} encoding,'' \emph{IEEE Trans. Inf. Theory}, vol.~68, no.~4,
  pp. 2704--2718, Apr. 2022.

\bibitem{MousaviMaddah-AliMirmohseni19_1}
M.~H. Mousavi, M.~A. Maddah-Ali, and M.~Mirmohseni, ``Private inner product
  retrieval for distributed machine learning,'' in \emph{Proc. IEEE Int. Symp.
  Inf. Theory (ISIT)}, Paris, France, Jul. 7--12, 2019, pp. 355--359.

\bibitem{ObeadLinRosnesKliewer19_2}
S.~A. Obead, H.-Y. Lin, E.~Rosnes, and J.~Kliewer, ``On the capacity of private
  nonlinear computation for replicated databases,'' in \emph{Proc. IEEE Inf.
  Theory Workshop (ITW)}, Visby, Sweden, Aug. 25--28, 2019.

\bibitem{ChenWangJafar20_1}
Z.~Chen, Z.~Wang, and S.~A. Jafar, ``The asymptotic capacity of private
  search,'' \emph{IEEE Trans. Inf. Theory}, vol.~66, no.~8, pp. 4709--4721,
  Aug. 2020.

\bibitem{Alon99_1}
N.~Alon, ``Combinatorial nullstellensatz,'' \emph{Combinatorics, Probability
  Comput.}, vol.~8, no. 1–2, pp. 7--29, Jan. 1999.

\bibitem{BondyMurty08_1}
J.~A. Bondy and U.~S.~R. Murty, \emph{Graph Theory}.\hskip 1em plus 0.5em minus
  0.4em\relax London, U.K.: Springer, 2008.

\bibitem{BehzadChartrandCooper67_1}
M.~Behzad, G.~Chartrand, and J.~K. {Cooper, Jr.}, ``The colour numbers of
  complete graphs,'' \emph{J. London Math. Soc.}, vol. s1-42, no.~1, pp.
  226--228, 1967.

\bibitem{Steinbach04_1}
P.~Steinbach, \emph{Field Guide to Simple Graphs}, 2nd~ed.\hskip 1em plus 0.5em
  minus 0.4em\relax Design Lab, 2004.

\bibitem{OEIS-A000088}
\BIBentryALTinterwordspacing
{OEIS Foundation Inc. (2024)}, \emph{Number of Graphs on $n$ Unlabeled Nodes.},
  {Entry A000088 in The On-Line Encyclopedia of Integer Sequences}. [Online].
  Available: \url{https://oeis.org/A000088}
\BIBentrySTDinterwordspacing

\bibitem{Koenig90_1}
D.~K{\"o}nig, \emph{Theory of Finite and Infinite Graphs}.\hskip 1em plus 0.5em
  minus 0.4em\relax Boston, USA: Birkh{\"a}user, 1990.

\bibitem{HagbergSchultSwart08_1}
A.~A. Hagberg, D.~A. Schult, and P.~J. Swart, ``Exploring network structure,
  dynamics, and function using {NetworkX},'' in \emph{Proc. 7th Python Sci.
  Conf. (SciPy)}, Pasadena, CA, USA, Aug. 19--24, 2008, pp. 11--15.

\end{thebibliography}
